\documentclass[journal]{IEEEtran}
\usepackage{graphicx}
\usepackage{cite}
\usepackage{amssymb,amsmath}
\usepackage{fancyhdr}

\usepackage{array}
\usepackage{color}
\usepackage{booktabs}
\usepackage{multirow}
\usepackage{subfig}
\usepackage{makecell}
\usepackage{pifont}
\usepackage{array}
\usepackage{amssymb}
\usepackage{booktabs}
\usepackage{doi}
\usepackage [latin1]{inputenc}
\ifCLASSINFOpdf
\else
\fi
\begin{document}
%\bibliographystyle{unsrt}
\title{
Differential Privacy for Industrial Internet of Things: Opportunities, Applications and Challenges}

         % <-this % stops a space

\author{Bin Jiang,~\IEEEmembership{Member,~IEEE},
        Jianqiang Li,
        Guanghui Yue,
       and Houbing Song,~\IEEEmembership{Senior Member,~IEEE}
         % <-this % stops a space
\thanks{Please cite the published version at IEEE Internet of Things Journal: https://doi.org/10.1109/JIOT.2021.3057419}
\thanks{This work was partially supported  by China
Postdoctoral Science Foundation (No. 2020M680125), National Natural Science Foundation
of China (No. U1713212) and the National Science Foundation under Grant No. 1956193. \emph{(Corresponding author: Jianqiang Li.)}}
\thanks{B. Jiang and J. Li are with College of Computer Science and Software Engineering, Shenzhen University, Shenzhen, 518000, China   (email: jiangbin@ieee.org; lijq@szu.edu.cn).}
\thanks{G. Yue is with School of Biomedical Engineering, Health Science Center, Shenzhen University, Shenzhen, 518000, China (email: yueguanghui@szu.edu.cn).}
\thanks{H. Song is with the Security and Optimization for Networked Globe Laboratory (SONG Lab), Department of Electrical Engineering and Computer Science, Embry-Riddle Aeronautical University, Daytona Beach, FL, 32114, USA. (email: h.song@ieee.org).}
}


\markboth{IEEE Internet of Things Journal}%
{Shell \MakeLowercase{\textit{et al.}}: Bare Demo of IEEEtran.cls for IEEE Journals}



\maketitle
\thispagestyle{fancy}
\fancyhead{}
\lhead{Accepted for publication by IEEE Internet of Things Journal, 2021,  DOI: 10.1109/JIOT.2021.3057419 }
\lfoot {\scriptsize{ \copyright~ 2021 IEEE. Personal use of this material is permitted. Permission from IEEE must be obtained for all other uses, in any current or future media, including reprinting/republishing this material for advertising or promotional purposes, creating new collective works, for resale or redistribution to servers or lists, or reuse of any copyrighted component of this work in other works.}}
\cfoot{}
\rfoot{}
\begin{abstract}

The development of Internet of Things (IoT) brings new changes to various fields. Particularly, industrial IoT (IIoT) is promoting a new round of industrial revolution. With more applications of IIoT, privacy protection issues are emerging. Especially, some common algorithms in IIoT technology, such as deep models, strongly rely on data collection, which leads to the risk of privacy disclosure. Recently, differential privacy has been used to protect user-terminal privacy in IIoT, so it is necessary to make in-depth research on this topic. In this article, we conduct a comprehensive survey on the opportunities, applications, and challenges of differential privacy in IIoT. We first review related papers on IIoT and privacy protection, respectively. Then, we focus on the metrics of industrial data privacy, and analyze the contradiction between data utilization for deep models and individual privacy protection. Several valuable problems are summarized and new research ideas are put forward. In conclusion, this survey is dedicated to complete comprehensive summary and lay foundation for the follow-up research on industrial differential privacy.

\end{abstract}


% \fi
\IEEEpeerreviewmaketitle

\section{Introduction}
\label{sec1}

\IEEEPARstart{T}{he rapid}  rise of Internet of Things (IoT) brings new demands and scenarios for humans daily life. For example, development of applications such as wearable devices \cite{Seneviratne2017survey}, smart appliances \cite{Ma2018Toward}, autonomous driving \cite{Wang2018Networking}, intelligent robots \cite{Morioka2004Human}, have prompted billions of new devices to connect by each other, which is accelerating interconnection in IoT system \cite{Tran2020Towards, song2016cyber}.
For industry, wireless communication and artificial intelligence (AI) jointly promote the development of industrial Internet of Things (IIoT) \cite{xu2017security}. Especially, IIoT continuously integrates various kinds of sensors and controllers with sensing network \cite{Chen2020Internet}, monitoring capabilities \cite{Zhao2019Design}, mobile communication \cite{Pereira2017Experimental}, intelligent analysis \cite{Shen2018Secure} and other technologies, so as to greatly improve manufacturing efficiency and product quality, reduce product cost and resource consumption, and finally achieve the upgrading of traditional industry \cite{Cai2020APrivate,Chong2018Learning}. In addition, a large number of industrial data are analyzed by cloud computing mode, so IIoT is essentially machine to machine (M2M) support that extends to the cloud and edge \cite{Shi2016Edge}.

Rapid development brings unexpected problems. Under the background of increasing application types, how to protect industrial individual privacy has become an important topic in IIoT \cite{Siddula2018Privacy,Pokhrel2020QoS, song2017security}.  In current research, various privacy protection methods have been applied to  IIoT technologies, and it has witnessed some effective algorithms \cite{zheng2018data, Lin2017A, Drosatos2020Privacy}. Among existing technologies, differential privacy has been identified as the most attractive, especially in the process of individual data publishing for the group network. For IIoT, the application of differential privacy is still in its infancy and trial stage, but this topic is very valuable and there are also preliminary research results to be summarized and compared \cite{Arachchige2020A}.

\begin{table*}
\centering
\caption{Related survey papers}
\begin{tabular}{|m{0.3cm}|m{0.4cm}| m{3cm}|m{6.3cm}|m{1cm}|m{1cm}|m{3cm}|}
\hline
\textbf{Ref} & \textbf{Year} & \makecell[c]{\textbf{Survey Description}}  & \makecell[c]{\textbf{Main Contributions}} & \textbf{DP Related} & \textbf{IIoT  Related} & \textbf{Distinction with This Survey}\\
\hline
\cite{2017Differentially} & 2017& Differentially private data publishing and analysis &  \makecell{Focus on practical aspects for differential privacy\\  Present the concepts and practical aspects of DP} & \ding{73}\ding{73}\ding{73}\ding{73}\ding{73} & \ding{73} &  It restricted observations to
data publishing and analysis scenarios\\
\hline
\cite{2017Survey} & 2017 & Improving data utility in differentially private sequential data publishing & \makecell{Classifications for existing DP-based methods \\ Utility comparison in different categories \\ Application scenario analysis } & \ding{73}\ding{73}\ding{73}\ding{73}\ding{73} & \ding{73}  &   It mainly considered differential privacy in  sequential data publishing\\
\hline
\cite{2018Industrial} & 2018 & Challenges, opportunities, and directions  in IIoT  & \makecell{Clarify the concepts of IoT, IIoT
and  Industry 4.0 \\  Highlight the
opportunities brought in by IIoT \\ Systematic overview of the state-of-the-art research } &\ding{73} & \ding{73}\ding{73}\ding{73}\ding{73}\ding{73}  & It focused  on the Internet of things as a whole, less content for privacy  \\
\hline
\cite{Wang2019Privacy} & 2018 & Privacy preservation in big data from the communication perspective & \makecell{Review privacy-preserving framework in big data \\ Especially differential privacy for big data
 \\ Consider it from communication perspective} & \ding{73}\ding{73}\ding{73} & \ding{73}\ding{73} & There is obvious distinction between big data and Industrial IoT \\
\hline
\cite{2020Differential} &  2019 & Differential privacy in cyber physical systems & \makecell{ Firstly highlight some
DP in CPSs domains \\ Outline certain open issues and challenges} & \ding{73}\ding{73}\ding{73}\ding{73}\ding{73} &  \ding{73}\ding{73}  & IIoT and CPS have obvious differences in application scenarios\\
\hline
\cite{Serror2020Challenges} &  2020 & Challenges and opportunities in securing the industrial Internet of things & \makecell{Identify
differences regarding security from IoT to IIoT \\  Derive distinct security goals and challenges in IIoT  \\ Survey current best practices for IIoT security } & \ding{73} &  \ding{73}\ding{73}\ding{73}\ding{73}\ding{73}  &  It considered most of the security issues for IIoT\\
\hline
\end{tabular}
\label{table01}
\end{table*}

\subsection{Motivations}

Many cases of privacy leakage in IIoT have been reported. For example, in industrial power consumption, an adversary can infer production efficiency and production type from different types of electricity consumption and peak power consumption period. In this context, the leakage of power consumption information will lead to more dangerous privacy issues. In this way, differential privacy in IIoT is very promising, and it is generally necessary to conduct further research in this direction. The motivations of this paper can be summarized in following three aspects.

Differential privacy in IIoT is not same as commonly used as in traditional IoT systems. Specially, the privacy protection of IIoT is an integrated problem. In order to design novel algorithms, it not only considers the inherent characteristics of IIoT, but also fully exploits the advanced combination of differential privacy with industry demand.  So it is the primary motivation of this paper, in order to present the state of the art of differential privacy in IIoT.

Big data and intelligent decision gradually play a dynamic role in the field of modern IIoT \cite{Cai2019Trading}, and the privacy protection issues in this process are gradually highlighted. We hope that this survey can help researchers to promote the power of differential privacy in IIoT, and it can also attract more researchers to pay attention to this topic. Especially, data distribution has become an inevitable choice for the development of IIoT.

While differential privacy in IIoT has been  recognized, many open issues need to be identified to provide guidance for the follow-up research, which is also an important motivation   in this paper. In second half of this survey, it will summarize some valuable research hotspots for reference. In this way, it can indirectly promote the development of IIoT privacy protection, in order to solve the existing important problems.

\subsection{Related Survey Papers}

 It is very important to collect and compare the survey papers in this topic or the papers which are related to privacy in IIoT. There are several survey papers as listed in Table \ref{table01}.

Zhu \emph{et al.} \cite{2017Differentially}  provided a structured survey on differentially private data publishing and analysis. Especially, authors discussed this issue along two directions: 1) data publishing and 2) data analysis on differential privacy. The typical algorithms are analyzed and compared for diverse mechanisms of  differential privacy.

 Yang \emph{et al.} \cite{2017Survey} investigated the existing schemes on differentially private sequential data publishing, from the perspective of proving data utility. Specially, authors summarized this topic from five aspects:   1) distribution optimization; 2) correlations exploitation; 3) sensitivity calibration; 4) transformation and 5) decomposition.

 Sisinni \emph{et al.} \cite{2018Industrial} conducted a comprehensive survey on IIoT, and authors clarified the concepts of IoT, IIoT, and Industry 4.0. In this article,  authors made an in-depth analysis on IIoT, and focused on the development of intelligent manufacturing (IM).

 Wang \emph{et al.} \cite{Wang2019Privacy}  systematically summarized the privacy preservation with the advancement in big data, and particularly from the  communication perspective. In this survey, authors discussed the related issues on sensitive information about individuals and covered the fundamental privacy-preserving framework, especially on differential privacy.

 Hassan \emph{et al.} \cite{2020Differential} presented a comprehensive survey on differential privacy in cyber-physical systems (CPS). Specially, authors summarized the application and implementation based on energy systems, transportation systems, healthcare and medical systems, and IIoT, which can be classified as four major applications.

Serror \emph{et al.} \cite{Serror2020Challenges} conducted a survey of security  issues in IIoT with challenges and opportunities for future reeaserch. In many scenarios, information security and data privacy can work together. Therefore, the summary work done by the authors also provides some important references for this field.

Compared with the above related reviews, this survey mainly focuses on differential privacy on IIoT, which is the obvious distinction with existing survey papers. The special comparisons can be referred in  Table \ref{table01}.

\begin{table}
\centering
\caption{Abbreviations in this paper}
\begin{tabular}{|c|c|}
\hline
\specialrule{0em}{1pt}{1pt}\textbf{Abbreviation} & \textbf{Referred}\\
\hline
DP & Differential privacy\\
\hline
IoT & Internet of things\\
\hline
IIoT & Industrial Internet of things\\
\hline
M2M & Machine to machine\\
\hline
IM & Intelligent manufacturing \\
\hline
LBS & Location based services\\
\hline
AMR & Autonomous mobile robot\\
\hline
CPS & Cyber physical systems\\
\hline
GAN & Generative adversarial networks \\
\hline
RNN & Recurrent neural network \\
\hline
PSPU & Pseudonym swap with provable unlinkability \\
\hline
UAV & Unmanned aerial vehicle\\
\hline
PLO & Power line obfuscation\\
\hline
LDP & Local differential privacy\\
\hline
SSA& Singular spectrum analysis \\
\hline
CDP& Cost-friendly differential privacy-preserving \\
\hline
OPF& Optimal power flow  \\
\hline
PPDP & Privacy-preserving data publishing  \\
\hline
DPLP& Differential privacy-based location protection  \\
\hline
RA-SP & Risk-averse two-stage stochastic problem\\
\hline
DRW & Directed random walk \\
\hline
STBD & Spatial temporal budget distribution \\
\hline
DAOs & Decentralized autonomous organizations \\
\hline
VANETs & Wireless communication, vehicle ad hoc networks \\
\hline
i2b2 & Informatics for integrating biology and the bedside \\
\hline
AGV & Automated guided vehicle \\
\hline
QRC & Quick response code\\
\hline
RFID & Radio frequency identification \\
\hline
ERP & Enterprise resource planning\\
\hline
\end{tabular}
\label{table02}
\end{table}

\begin{figure*}
\centering
\includegraphics[width=6in,height=3.3in]{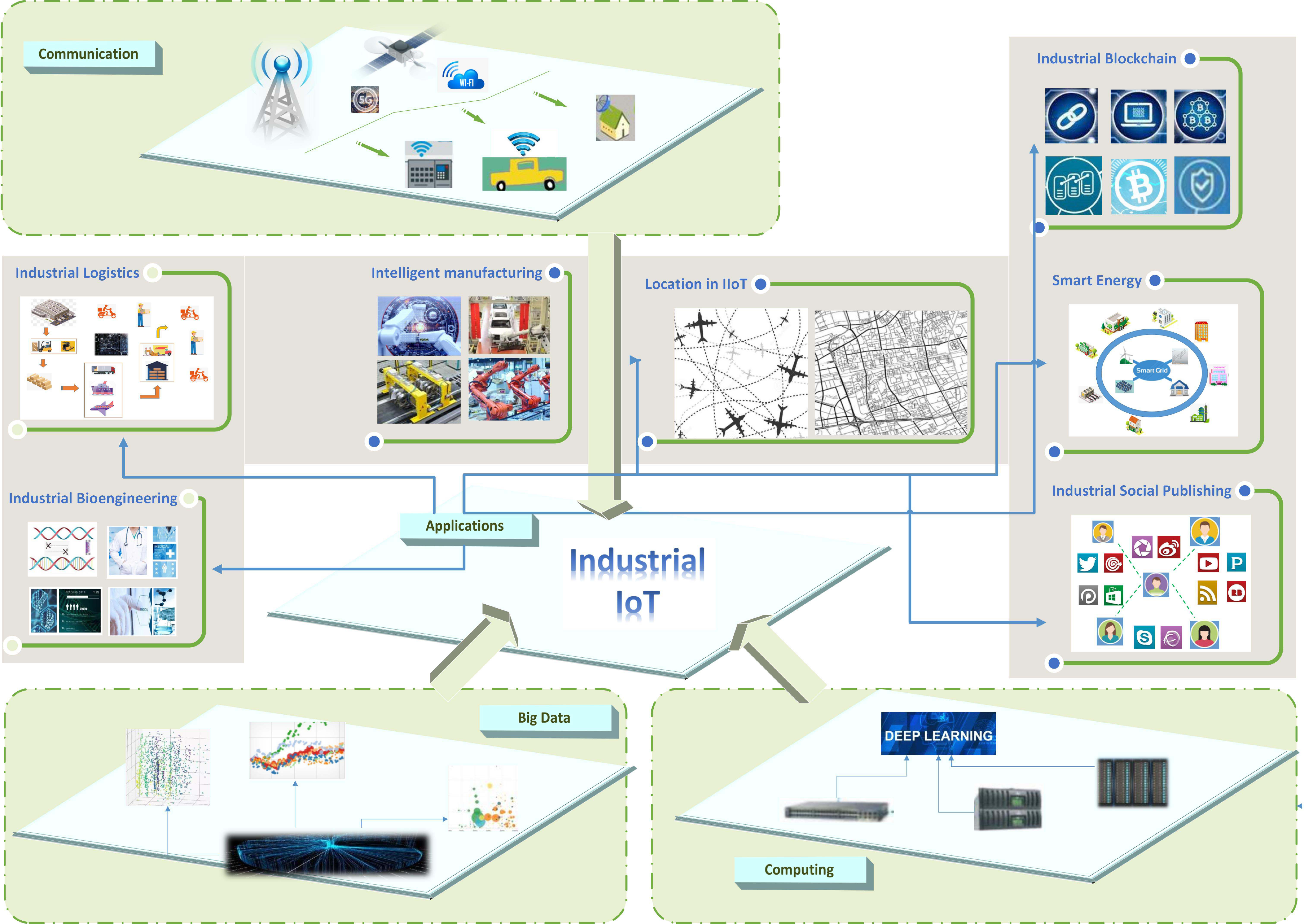}
\caption{The architecture of IIoT}
\label{figure01}
\end{figure*}

\subsection{Contributions and Organization}
As a survey paper, it aims to investigate the application value and potential of differential privacy in the field of IIoT. In this paper, it focuses on the following issues: privacy measurement in IIoT system, data contradiction between deep network and differential privacy, analysis on industrial applications of differential privacy,  existing problems and solutions in this field. On the whole, the main contributions of this paper can be summarized as follows:

1) As far as we know, there is no direct review article to summarize the application of differential privacy in IIoT. As a new research direction, this paper summarizes the related privacy issues in the IIoT, and compares the existing methods.

2) This paper discusses the privacy measurement and the balance of privacy data in deep models.  These two subtopics are very popular in the field of privacy protection. This paper summarizes its application migration in the field of IIoT.

3) In this paper, the application cases, existing problems and solutions of differential privacy in IIoT are analyzed in detail. For future researchers, it can lay the foundation and provide more valuable new ideas.

The abbreviations  used in this paper are presented in Table \ref{table02} and readers can refer to the abbreviations used in this paper. The remainder of this survey is organized as follows. In Sections \ref{sec2} and \ref{sec3}, we conduct brief surveys on IIoT and differential privacy, respectively. In Sections \ref{sec4} and \ref{sec5}, we discuss the two hot issues in this topic: privacy metrics and the conflicts between differential privacy and deep learning. In Section \ref{sec6}, we will present and compare some typical applications based on privacy in IIoT. More importantly, this paper will conclude the current problems and valuable issues in this topic and present future research directions, which will be included in  Sections \ref{sec7} and \ref{sec8}, respectively. At last, the final conclusions are given in Section \ref{sec9}.

\section{Privacy Protection in Industrial IoT}
\label{sec2}

\subsection{From IoT to IIoT}

The concept and development of IoT has been widely known \cite{Xu2014Internet}, while the definition of IIoT is often confused. In fact, the generalized IIoT contains  not only the common electronic equipment used in traditional IoT, but also achieves the connection of intelligent industrial production equipment \cite{2018Industrial}. In short, the concept of IIoT simply points to the industrial production process based on IoT. Generally, the architecture of IIoT can be shown in Fig. \ref{figure01}.

In IIoT, operators can connect devices, assets and sensors to collect undeveloped data. It also enables users to deliver scalable, reliable applications faster to meet the changing needs of industrial customers \cite{rawat2017industrial}. For example, in connected factory, sensor-enabled device can provide data information which can be analyzed to predict when and where the equipment will fail, thus helping the factory to prevent shutdown. If fault occurs, factory can analyze the data to identify the problem and take corrective actions to prevent it from occurring in advance \cite{Zhang2017An}.

Liang \emph{et al.} \cite{2020Toward}  summarized the edge computing and deep learning technology in IIoT. In addition, authors leveraged the edge computing paradigm and proposed an edge computing-based deep learning model, which can migrate deep learning between cloud servers and edge nodes.

In order to achieve defined constrained optimization, Dai  \emph{et al.} \cite{2019A}  proposed a nature-inspired genetic algorithm to keep connected confident information coverage, which is very important for IIoT. However, this issue is very specific, and it is not a general summary and analysis of IIoT.

In summary, the IIoT has three main characteristics. Firstly, IIoT mainly emphasizes the application of reproduction and service, usually involving higher value equipment and assets, such as energy, transportation and industrial control. At the same time, it also has higher requirements for data operation security \cite{Butun2019security}. On the contrary, the traditional IoT pays more attention to the field of consumption, such as home application. Secondly, IIoT is built on the industrial infrastructure and is used to upgrade rather than replace the original industrial production equipment. Thirdly, IIoT can be considered as a subset of the IoT, which focuses on productivity improvement.

\begin{table*}
\centering
\caption{Typical papers on security and privacy in IIoT}
\begin{tabular}{|m{2cm}|m{0.7cm}|m{4.5cm}|m{0.8cm}|m{3cm}|m{1.5cm}<{\centering}|}
\hline
\specialrule{0em}{2pt}{2pt}\textbf{Authors} &\textbf{Ref} & \makecell[c]{\textbf{Description}} & \textbf{Year} & \textbf{Focused Subfield in IIoT} & \textbf{On Privacy} \\
\hline
Mouratidis  \emph{et al.}  &\cite{2018A} & Security analysis method for industrial Internet of things & 2018& Security analysis & \checkmark\\
\hline
Dai  \emph{et al.}  &\cite{2019A} & Nature-inspired node deployment strategy for connected confident IIoT & 2019 & Connected confidence & $\times$\\
\hline
Mcginthy \emph{et al.} &\cite{2019Secure} & Critical infrastructure node design for secure industrial Internet of things  & 2019 & Infrastructure node design& $\times$\\
\hline
Al-Hawawreh  \emph{et al.} &\cite{2019Targeted} & Targeted ransomware in edge system of brownfield industrial Internet of things & 2019& Edge system & $\times$\\
\hline
Urbina  \emph{et al.}  & \cite{2019Smart}  & SoC architecture that satisfies IIoT operational requirements & 2019&  Smart sensor & $\times$\\
\hline
Zolanvari \emph{et al.}  &\cite{2019Machine} & Deep learning driven network vulnerability analysis for IIoT & 2019& Network vulnerability  & \checkmark\\
\hline
Wang \emph{et al.} &\cite{2020MTES} &  Trust evaluation scheme in sensor-cloud-enabled industrial Internet of things & 2019 & Trust evaluation scheme & \checkmark\\
\hline
Boudagdigue \emph{et al.} &\cite{2020Trust} & Dynamic trust management model suitable for industrial environments & 2020 &Trust management & \checkmark\\
\hline
Zheng \emph{et al.}  &\cite{Zheng2020p} & Privacy-preserved data sharing towards multiple parties in Industrial IoTs & 2020&  Multiple privacy & \checkmark\\
\hline
\end{tabular}
\label{table03}
\end{table*}

\subsection{IIoT Characteristics with Main Technologies}
Before the emergence of the IIoT, many industrial scenarios still did not have networking capacity, or only provided one-way communication. With IIoT, two-way communication can be achieved: 1) data is provided to the controller and cloud, and 2) feedback is provided to the terminal. For example, a production run can be supported by changing the parameters on the sensor. IIoT provides the opportunity to collect and utilize previously unused information from the warehouse to the plant floor, and correlate existing and new different data sets, ultimately driving improvement and forming new solutions. In this survey, we summarize the main characteristics of IIoT in five aspects: 1) intelligent self-decision; 2) real-time monitoring; 3) smart operation; 4) logistics optimization and 5) energy control.

\subsubsection{Intelligent Self-decision}

For industrial production, the improvement of equipment self decision-making ability means the basic realization of IIoT. In this mode, industrial equipment can make self decision and adjust the production process under certain authority \cite{Tang2020A}. For example, it is very difficult for any equipment to produce products or provide services with full rate. Through the intelligent  self-decision, it can automatically detect the unqualified products, so as to eliminate the low-quality products.

\subsubsection{Real-time Monitoring}

After the realization of IIoT, the supervision of industrial production will become easier. Real-time monitoring and error correction has become an important guarantee for industrial products and services \cite{Zhao2019Design}. For normal industrial operation process, it is lack of response to emergency. Through real-time monitoring in IIoT, it can predict and evaluate the possible risks in advance and effectively eliminate.

\subsubsection{Smart Operation}

For industrial products and services, the operation efficiency is very important. Through smart operation based on IIoT, it can  improve productivity and achieve valid operation \cite{Fortino2010trust}. At the same time, smart operation in IIoT can be competent for more complex and dangerous production activities, which can be in  low-risk.

\subsubsection{Logistics Optimization}

In the context of IIoT, intelligent logistics is very important. IIoT  can also make it more intelligent, automated warehouse management, including warehousing, inventory, outbound, picking, replenishment, delivery and so on, which can ensure  timely and accurately grasping real inventory data \cite{Guo0CPS}.  Intelligent logistics is the use of integrated intelligent technology, so that the logistics system can imitate human intelligence, with the ability of thinking, perception, learning, reasoning and judgment and solving some problems in logistics by itself.

\subsubsection{Smart Energy}
IIoT provides better energy management solutions for production, which can not only effectively help  improve production efficiency, but also maximize energy utilization, so as to reduce unnecessary waste and environmental pollution, and help save energy and increase production \cite{Sun2016A}.

\subsection{Industrial Privacy}

In the promotion process of IIoT, it is also constantly putting forward new requirements and application scenarios, especially on privacy. From the application perspective, there are many aspects on privacy issue in IIoT. In this paper, we summarized it into three main categories: 1) network security; 2) data value and 3) interconnection protocol.

\subsubsection{Network Security}

IIoT is a complex infrastructure involving Internet and data transmission. Therefore, it has unprecedented network attack potential. The ownership and security of data are the main challenges of IIoT applications \cite{Ma2019A}. In addition, the bad performance of network security brings more privacy disclosing.

\subsubsection{Data Value}
In general, IIoT data consider how to collect more data and how to process data, but  how to find the value of data is more important.  Data is an important part of IIoT system, and user privacy data is usually the most critical factor. Therefore, how to mine the potential value of user information is related to effectiveness and efficiency of IIoT \cite{Luo2015Efficient}.

\subsubsection{Interconnection Protocol}
IoT is the foundation of the IIoT, but it is not enough to connect the sensor with the equipment, but also need to connect with the Internet. However, for the wireless device network with relatively short distance, there are several protocols competing locally, such as Bluetooth, ZigBee and thread, which must face the interoperability problem \cite{Hassan2020An}. Privacy risk will be brought by the complex interconnection protocol.

For the IIoT, the protection of user privacy has been paid more and more attentions. Therefore, the privacy issues in the IIoT can be classified and summarized in Table \ref{table03}.

 Mouratidis   and Diamantopoulou \cite{2018A} conducted an analysis on security requirements and identification of attack paths for mitigation of potential vulnerabilities. Mcginthy \emph{et al.}  \cite{2019Secure} put forward the critical infrastructure node design for IIoT. Al-Hawawreh  \emph{et al.} \cite{2019Targeted} analyzed the cyber threat called as targeted ransomware for fog-computing IIoT.  Urbina  \emph{et al.} \cite{2019Smart} designed the SoC architecture defined as smart sensor for IIoT.

Zolanvari \emph{et al.} \cite{2019Machine}  made a network vulnerability analysis based on deep models for IIoT. Wang \emph{et al.} \cite{2020MTES} proposed an intelligent trust evaluation method for IIoT considering sensor-cloud-enabled condition . Boudagdigue \emph{et al.}  \cite{2020Trust} discussed the trust management issues in IIoT . Zheng and Cai \cite{Zheng2020p}   designed a privacy-preserved data sharing method which can be used for multiple parties in IIoT.

All the above researches have investigated the privacy security of IIoT from different perspectives. However, the use of differential privacy technology is still in its infancy, and it is only a preliminary attempt. Therefore, it is necessary to explore the relationship between differential privacy and IoT.

\section{Development and Opportunities of Differential Privacy for Industrial IoT}
\label{sec3}
\begin{figure}
\centering
\includegraphics[width=3.2in,height=2.2in]{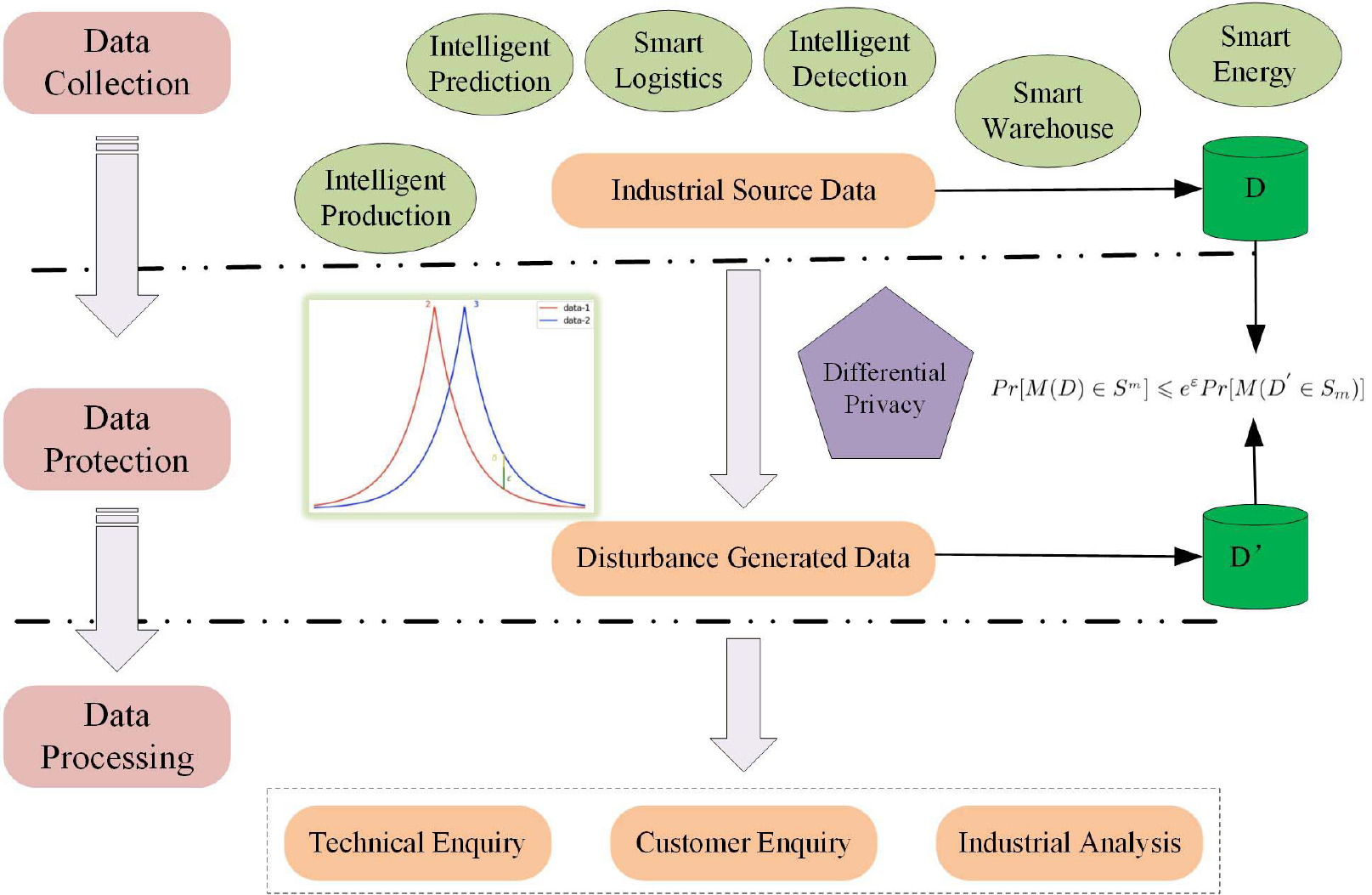}
\caption{Differential privacy in IIoT}
\label{figure02}
\end{figure}

Differential privacy  is widely accepted as a strict privacy protection model. Before the advent of differential privacy, the existing privacy preserving algorithms are still problematic, such as k-anonymity \cite{wang2019achieving}. And differential privacy uses more stringent constraints and definitions. It protects the potential user privacy information in the published data by adding interference noise. Even if the attacker has mastered certain information, it still can't infer the information. Therefore, this is a method to completely eliminate the possibility of privacy information disclosure from the data source and the detailed technology flow is shown in Fig. \ref{figure02}.

The design goal of differential privacy is to complete the analysis of the whole data set without disclosing the information of a single sample. On the one hand, differential privacy can resist the attacker's possible background knowledge. On the other hand, differential privacy is based on a solid mathematical foundation and can quantitatively evaluate the memory of privacy protection.

It is well known that Dwork and Roth \cite{Dwork2013The} proposed differential privacy and authors explained the related algorithmic foundations of differential privacy. Li \emph{et al.} \cite{Li2016Differential} analyzed differential privacy from theory to practice and relevant contents are summarized in depth.

\subsection{Definitions}

It is necessary to give the most basic definition of travel privacy here. For a random algorithm $M$, $P_{m}$ is the set of all the values that algorithm $M$ can output. If for any pair of adjacent data sets $D$ and $D^{'}$, any subset $S_{m}$ of $P_{m}$, algorithm $M$ satisfies

\begin{equation}
Pr[M(D) \in S_{m}]\leqslant e^{\varepsilon  }Pr[M(D^{'})\in S_{m}]
\label{eq1}
\end{equation}

Then the algorithm $M$ satisfies differential privacy, where $\varepsilon$ is the privacy protection budget.

For the definition of differential privacy,  researchers are also fully mining, exploring and expanding.
Pathak \emph{et al.}  \cite{2012Large} proposed the large margin gaussian mixture models based on differential privacy. In their work, the greatest contribution is the large marginal loss function with perturbed regularities.  Kairouz \emph{et al.}  \cite{2015The}  discussed the optimal multidimensional setting mechanism in differential privacy, which improved the optimal mechanism while protecting privacy. Wang \emph{et al.} \cite{Wang2016On}  analyzed the relationship between identifiability, mutual-information privacy and differential privacy.  Inan \emph{et al.} \cite{2017Sensitivity}  discussed the sensitivity analysis for non-interactive differential privacy. They studied how to answer  statistical range queries with batch mode accurately in privacy condition. For caching problem in IoT, Zhang \emph{et al.}  proposed a data-driven caching method for information-centric networks based on local differential privacy in \cite{Zhang2018Data}. In order to solve the multi-party data publishing, Cheng \emph{et al.} \cite{2019Multi}  proposed the multi-party publishing method for high-dimensional data in differential privacy. Oneto \emph{et al.} \cite{Oneto2020Randomized}  addressed randomized learning and generalization for private classifiers. Specially, authors considered the problem from PAC-Bayes to stability and imported the definition of differential privacy.

\subsection{Progress and Development}

At the same time, many methods extended by differential privacy are proposed. Of course, these developments are usually produced in the context of changes in information technology. For example, the popularity of the IoT puts forward new requirements for the privacy protection of sensor data, and the emerging research is constantly approaching this direction.
In this survey, we summarize the relevant information and preliminary work, and summarize the development of differential privacy.

Xiao \emph{et al.} \cite{Xiao2011Differential} tried  differential privacy based on wavelet transforms and achieved better results.   Fouad \emph{et al.} \cite{Fouad2014A} proposed the supermodularity-based differential privacy method, which is valid on data anonymization. For individualdifferential privacy, Soria-Comas \emph{et al.}  proposed the utility-preserving formulation for  differential privacy guarantees in \cite{2016Individual}.

Wang \emph{et al.} \cite{Wang2017CTS}  developed the CTS-DP, which can be used for  correlated time-series data publishing based on differential privacy. Goryczka and Xiong \cite{2017A}  made  comprehensive comparison on multiparty secure additions while implementing differential privacy.  At the same time, Zhang \emph{et al.}  \cite{Zhu2017Dynamic} solved the ADMM-based distributed classification learning based on dynamic differential privacy. In addition, Cao \emph{et al.} quantified the differential privacy for continuous data release with temporal correlations in \cite{2017Quantifying}.

Kalantari \emph{et al. }  \cite{Kalantari2018Robust}  achieved the  robust privacy-utility tradeoffs with the combination of hamming distortion and  differential privacy.
Du \emph{et al.} \cite{Du2020Differential} designed the training model for differential privacy under the condition of wireless big data and  fog computing. In addition, Ou \emph{et al.} proposed the releasing correlated trajectories  in \cite{2018Releasing}, which can be used  with high utility and optimal differential privacy.

Zhang \emph{et al.}  \cite{2019Correlated}  proposed the  correlated differential privacy and it is helpful for feature selection in machine learning. Brinkrolf \emph{et al.} \cite{Brinkrolf2019Differential}  put forward the differential privacy method for learning vector quantization. In addition, Gong \emph{et al.}  made use of differential privacy for regression analysis based on relevance in \cite{Gong2019Differential}, and Ke \emph{et al.}   proposed the AQ-DP in \cite{Ke2019AQ}, which can be used as  new differential privacy scheme designed for big data quasi-identifier classifying. Huang \emph{et al.} \cite{2019An} put forward a method for  logistic classification mechanism based on differential privacy.

Li \emph{et al.} \cite{Li2020Secure} designed the secure metric learning method based on differential pairwise privacy.  Katewa \emph{et al.} \cite{2020Differen}  conducted a survey on differential privacy in network identification.

\subsection{Implementation Mechanisms}

Implementing differential privacy is very important, and adding noise is the main technology to achieve differential privacy protection. As two common addition mechanisms, Laplace mechanism is suitable for numerical results, while exponential mechanism is suitable for non numerical results \cite{2016The}. Some typical studies have also discussed the mechanism of noise addition.

Soria-Comas and Domingo-Ferrer \cite{Soria2013Optimal} proposed the optimal data-independent noise method for differential privacy implementation. For approximate differential privacy, Geng \emph{et al.} discussed the related optimal noise adding method in \cite{2016Optimal}.  In addition, Wang and Xu  analyzed the principal component for local differential privacy in \cite{Wang2019Principal}. From the perspective of engineering technology, the mainstream implementation schemes of differential privacy are generally classified into two types: 1)laplace mechanism and 2) index mechanism.

\subsubsection{Laplace Mechanism}

% https://blog.csdn.net/gosantiago/article/details/96139475?utm_medium=distribute.pc_relevant.none-task-blog-title-2&spm=1001.2101.3001.4242
% https://blog.csdn.net/maththinker/article/details/51464273

Given the data set $D$, the sensitivity of function $f:D\rightarrow R^{d}$ is defined as $\Delta f$, then the random algorithm  $M(D)=f(D)+Y$  provides differential privacy protection.
And $Y\rightarrow Lap(\Delta f/\varepsilon )$ is the random noise, which obeys the Laplace distribution with the scale parameter $\Delta f/\varepsilon $.

\begin{equation}
M(D)=f(D)+(Lap_{1}(\frac{\Delta f}{\varepsilon }),Lap_{2}(\frac{\Delta f}{\varepsilon }),...,Lap_{d}(\frac{\Delta f}{\varepsilon }))^{T}
\end{equation}

Li \emph{et al.} \cite{Li2019The}  discussed the optimal upper bound of the number of queries for Laplace mechanism under differential privacy.

\subsubsection{Index Mechanism}
The input of algorithm $M$ is data set $D$, and the output is an entity object $r\in Range$, $q(D,r)$ is an availability function, $\Delta q$ is the  sensitivity of function $q(D,r)$. If the algorithm $M$  is proportional to the probability of $exp(\frac{\varepsilon q(D,r)}{2\Delta q})$, all possible values can be normalized to get the corresponding probability value, then it can provide differential privacy protection.

\subsection{Basic Properties}

As far as differential privacy is concerned, it has many inherent properties, which can be used directly by algorithm designers.

For inherent properties, Geng  discussed in depth and improves the differential privacy mechanism by using the related properties. In \cite{geng2014optimal} and \cite{Geng2015The}, they achieved the optimal mechanism in differential privacy with help of multidimensional setting. Niu \emph{et al.} \cite{Niu2020Utility} also proposed the staircase mechanism for differential privacy.

After literature review, we summarize four properties of differential privacy as follows: 1) sequential synthesis; 2) parallel synthesis; 3) transformation invariance and 4) convexity.

\subsubsection{Sequential Synthesis}
If a differential privacy algorithm is composed of several algorithms, the level of privacy protection of the composite algorithm is the total budget of all algorithms. And this property is conditionally for the same dataset.

\subsubsection{Parallel Synthesis}
Suppose there are algorithms $M_{1},M_{2},...,M_{n}$, and the budget for privacy protection is $\varepsilon _{1},\varepsilon_{2},...,\varepsilon_{n}$. Then, for disjoint dataset $D _{1},D_{2},...,D_{n}$, the combination algorithm $M(M_{1}(D),M_{2}(D),...,M_{n}(D))$ composed of these algorithms provides differential privacy protection.

\subsubsection{Transformation Invariance}

Given that any algorithm $A_{1}$ satisfies differential privacy, $A(\cdot )=A_{2}(A_{1}(\cdot))$ satisfies differential privacy for any algorithm $A_{2}$, where $A_{2}$ does not necessarily satisfy differential privacy.

\subsubsection{Convexity}

Given two algorithms $A_{1}$ and $A_{2}$, both of them satisfy differential privacy. For any probability $p\in [0,1]$, the symbol $A_{p}$ is used as a mechanism, which uses the probability of $p$ to use algorithm $A_{1}$ and the probability of $1-p$ to use algorithm $A_{2}$, then mechanism $A_{p}$ satisfies differential privacy.

Based on the above properties and definitions, differential privacy algorithm is constantly improved to meet the needs of IIoT environment. The demand of differential privacy of IIoT is gradually rising, which has experienced the transformation from traditional IoT to IIoT \cite{Xu2018Distilling}.

\section{Metrics for Differential Privacy in IIoT}
\label{sec4}

Differential privacy is an important tool for privacy protection in the field of data publishing, but its advantages and disadvantages can only be evaluated posteriorly, and it is highly dependent on the privacy budget of empirical choice.

Some previous researches have been done for privacy metrics.  Chen \emph{et al.} \cite{2017Evaluating}  proposed an evaluation method on the risk of data disclosure under differential privacy and it is based on noise estimation.   Zhao and Wagner \cite{Zhao2019On}  discussed the privacy metrics on vehicular communication technologies and authors compared 41 privacy metrics in terms of four novel criteria.

Traditional privacy algorithm evaluation relies on the quantization of probability. Specifically, subjective probability judgment is widely used. However, the IIoT under the background of big data puts forward dynamic requirements for privacy risk assessment \cite{Xiong2018Enhancing}. In this survey, we summarize the current state of the art in privacy metrics for IIoT, especially differential privacy.

For privacy measurement, it mainly considers the choice of privacy protection technology and the professional background of attackers. In this way, privacy measurement originated from privacy anonymity technology \cite{Rassouli2019Optimal}.
When the risk of privacy disclosure in data is zero, the data achieves perfect privacy protection, which can achieve the maximum protection of privacy information. The data without any protection measures is regarded with the greatest risk of privacy information disclosure. Previous researches have proposed privacy measurement methods based on information entropy, set pair theory and differential privacy. The traditional privacy measurement methods mainly aim at small-scale and structured data stored in traditional relational databases\cite{Gunlu2018Privacy}.

In addition, there are also some methods for privacy measurement on cloud data. While cloud data are usually in industrial level, large-scale, multi-source, with multi-dimensional and unstructured mode. Therefore, compared with the traditional privacy protection measurement technology, the privacy measurement of cloud data should consider not only the privacy leakage measurement of small-scale and structured data, but also the privacy leakage measurement of large-scale and unstructured data.

 This survey will focus on privacy protection principle, measurement effect and main advantages and disadvantages. Generally, we summarize the existing privacy measurement for differential privacy in IIoT.

\subsection{Graph Theory and Mutual Information}

Differential privacy quantification model based on graph theory and mutual information can be used for  privacy leakage calculation method.  Based on the distance regularization and point transfer of graph, the privacy disclosure mutual information quantification method can be used, and the information upper bound of differential privacy disclosure has been proved and calculated \cite{2020Differential}.

The analysis and comparisons have shown that the privacy disclosure upper bound has a good functional relationship with the number of attributes, attribute values and privacy budget parameters of the original data set. Graph theory and mutual information can provide theoretical support for the design and evaluation of differential privacy algorithms.

\subsection{Information Entropy Measurement}

Aiming at the problem of composite data set with non interactive multi-attribute based on differential privacy, the information entropy measurement can be used to build assessment method for privacy degree, data utility and privacy leakage risk.

In principle, mutual information is used to analyze the attribute correlation, and the relational dependency graph model is used to express the attribute \cite{Naghibi2014A}. Based on the key privacy leakage paths in the graph, Markov privacy disclosure chain is constructed, and an associated attribute privacy measurement model and method can be proposed based on information entropy, which can effectively measure the amount of privacy leakage caused by associated attributes \cite{Zhu2020more}.

\subsection{Utility Optimization with Rate Distortion}

The solution of the contradiction between privacy protection and data utility is a research hotspot in the field of privacy protection, which will be discussed in Section \ref{sec5}. Aiming at the problem of privacy and utility balance in the off-line data publishing scenario of differential privacy, the optimal differential privacy mechanism to balance privacy and data utility is studied by using rate distortion theory \cite{2020Differential}. Based on the communication theory, such method abstracts the noise channel model of differential privacy, measures the privacy and utility of data publishing by mutual information and distortion function, and constructs the optimization model based on rate distortion theory \cite{2017Evaluating}.

\section{Conflicts: Differential Privacy and Deep Models}
\label{sec5}

In the application of IIoT, AI methods driven by deep learning are playing an important role. The performance of deep models strongly depends on the data size and data quality. Meanwhile, the widespread use of data brings the risk of privacy disclosure. Therefore, the contradiction between the two is often concerned by researchers. In order to establish a more stable and secure IIoT, it is necessary to consider how to improve data availability and fully protect privacy. Ha \emph{et al.} \cite{Ha2020Differential} conducted a survey paper on differential privacy in deep learning, which discussed the conflicts between privacy protection and deep models.

\subsection{Relationship between Differential Privacy and Deep Learning}

The primary task of deep learning is data training and test. However, in the process of data collection, privacy disclosure will occur, which is not conducive to the development of AI.

It is well known that three changes have driven the great success of deep learning in various fields:
(1) Exponential increase of data volume: in the implementation of IIoT, it will be more convenient to collect data and establish certain scale of dataset for deep learning algorithm.
(2) Breakthrough in computing power: The increasing computing hardware such as  GPU clusters makes it available for deep learning in IIoT.
(3) Algorithm breakthrough: algorithm breakthrough for IIoT promotes AI technology maturity and practicality in industry.

Although deep learning brings great benefits, there is a need to collect a large number of data, which involves the industrial privacy information. The leakage of these privacy data will lead to unpredictable  problems for industrial operators. In view of this problem, many scholars have carried on thorough research.

Xu \emph{et al.} \cite{2019GANobfuscator}  proposed the GANobfuscator, which can deal with the mitigating information leakage while using generative adversarial networks (GAN) based on Differential Privacy. In addition, Zhu \emph{et al.} \cite{Zhu2020more}   made a discussion on  differential privacy in  AI, especially on multi-agent systems, reinforcement learning, and knowledge transfer.

\subsection{Attack Types in Industrial Deep Learning Model}

Chen \emph{et al.} \cite{Chen2020RNN}  proposed the RNN-DP, which can be used as a differential privacy scheme base on recurrent neural network (RNN). Based on this scheme, we can achieve dynamic trajectory privacy protection.

Gong \emph{et al.} \cite{Gong2020Preserving}  put forward the differential privacy based on adaptive noise imposition, which can be used for general deep neural networks.

It is necessary to summarize current attach types in deep learning, especially in IIoT. Generally, we can classify it into adversarial attacks and cooperative attacks.

\subsubsection{Adversarial Attacks}

In the field of machine learning, the deep model used for classification is usually easily affected by the adverse examples. At the same time, the learning model components have linear characteristics. In this context, attackers can construct adverse examples to achieve the purpose of attack.

\subsubsection{Cooperative Attacks}

In deep learning, the training set with large amount of data can get more accurate prediction model. However, data sets are usually unbalanced. In order to solve this problem, cooperation and complementarity is a common method. Different data providers expand the training set by sharing data. In this context, individual data providers will shield their private data, which will bring the disadvantage of data closeness.

\subsection{Federated Learning}

Federated learning is a popular privacy protection model in deep learning, and its performance is also worthy of attention. It is quite different from the privacy protection theories commonly used in big data and data mining, such as K anonymity and L diversity. Federated learning protects user data privacy through parameter exchange under special mechanism with homomorphic encryption. In this way, the data and model in federated learning will not be transmitted, so there is no possibility of leakage at the data level, and it does not violate more stringent data protection laws.

Wei \emph{et al. } \cite{2019Federated} proposed the NbAFL, which can be regarded as a framework on federated learning with differential privacy. Authors added artificial noises to the parameters at the clients side before aggregating. Nuria \emph{et al.} \cite{Nuria2020Federated} discussed the federated learning in differential privacy in view of software tools analysis. In addition, Hu \emph{et al.} \cite{Hu0Personalized} also put forward a method on  personalized federated learning with differential privacy.

The methods based on differential privacy commonly add noise to the data, or use generalization method to blur some sensitive attributes until the third party can not distinguish individuals, so that the data can not be restored with high probability, so as to protect privacy. However, in essence, these methods still carry out the transmission of original data, and there is a potential possibility of being attacked, and under the more stringent data protection schemes. Correspondingly, federated learning is a more powerful solution.

In addition, Lu \emph{et al.} \cite{Lu2019Privacy}   put forward the differentially private asynchronous federated learning towards mobile edge computing in urban environment, and Hao \emph{et al.} \cite{Hao2020Efficient}  proposed the efficient and privacy-enhanced federated learning for IIoT based on  AI.

\subsection{Final Balance}
The balance between differential privacy and deep learning model is very important. It mainly protects privacy in training and testing stages. In general, the defender can introduce noise for differential privacy, but it will reduce the accuracy of the model. How to balance the two has become a hot topic in academia.

Sarwate and Chaudhuri \cite{Sarwate2013Signal}  conducted a review paper on  machine learning and differential privacy in the view of signal processing, especially for continuous data. Authors discussed the topic on  algorithms design and current challenges. Wang \emph{et al.} \cite{2020DNN}  proposed the  DNN-DP which can be used for sensitive crowdsourcing data, considering the balance between  differential privacy and deep neural network. Zheng \emph{et al.} \cite{Zheng2020Preserving} discussed the balance problem between local differential privacy and federated machine learning .

\section{Applications of Differential Privacy in IIoT}
\label{sec6}

\begin{figure*}
\centering
\includegraphics[width=7in,height=4.7in]{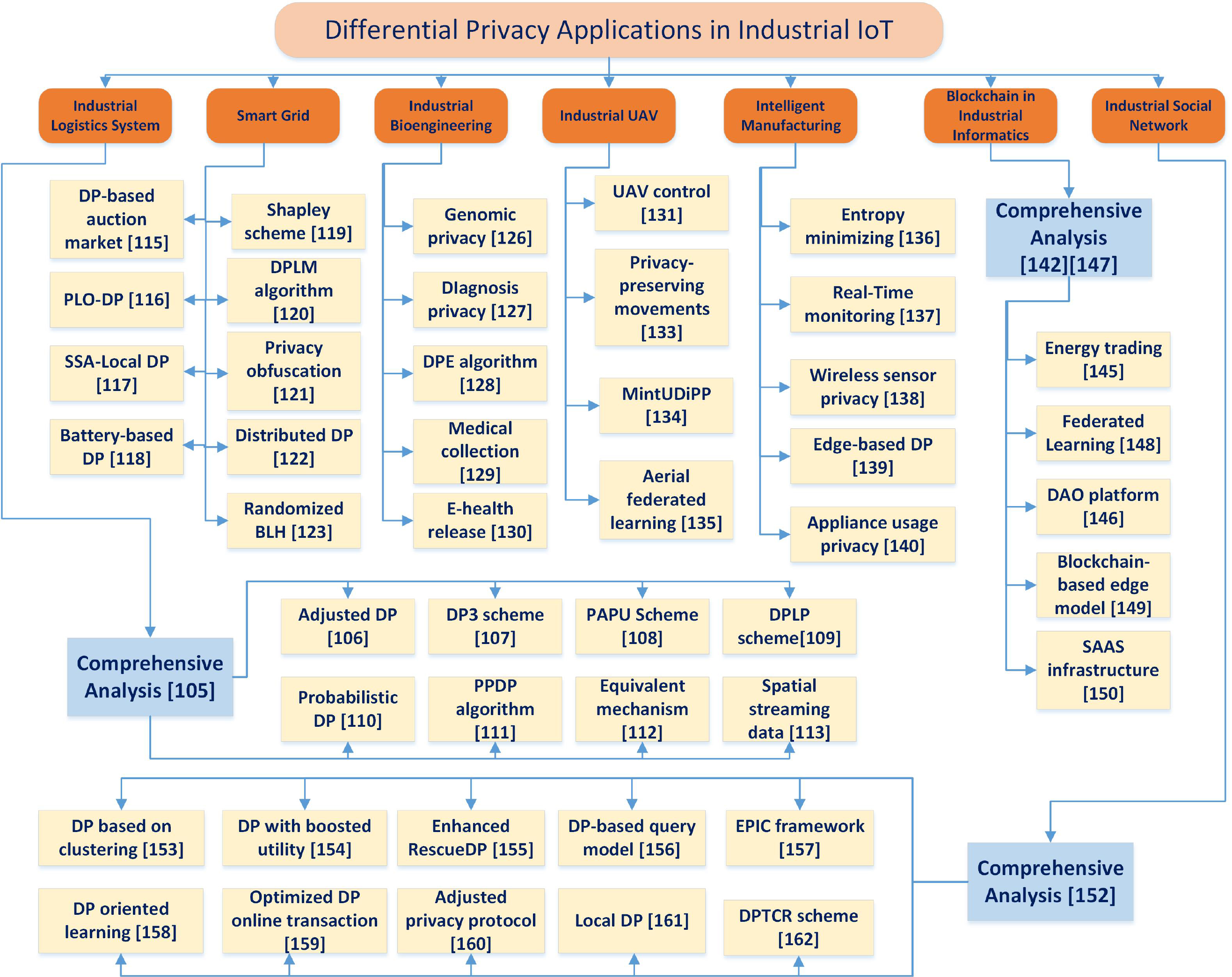}
\caption{Industrial application scenarios based on differential privacy}
\label{application}
\end{figure*}

Traditional industrial network modes are unable to meet the requirements of modern industry in terms of computing power, interaction speed, data analysis and so on.
After years of development, the concept of new IIoT has been widely understood and accepted by industry.
It is worth discussing what improvements differential privacy have been made to industrial scenarios.

Generally, the core elements of IIoT contain intelligent machine, advanced analytic and human-machine interaction. In this survey, we summarize the application scenarios in following aspects: 1) industrial logistics system; 2) smart grid; 3) industrial bioengineering; 4) industrial unmanned aerial vehicle; 5) intelligent manufacturing; 6) industrial blockchain and 7)industrial social network. And the whole review flow can be found in Fig. \ref{application}.

\begin{figure}
\centering
\includegraphics[width=2.5in,height=2.1in]{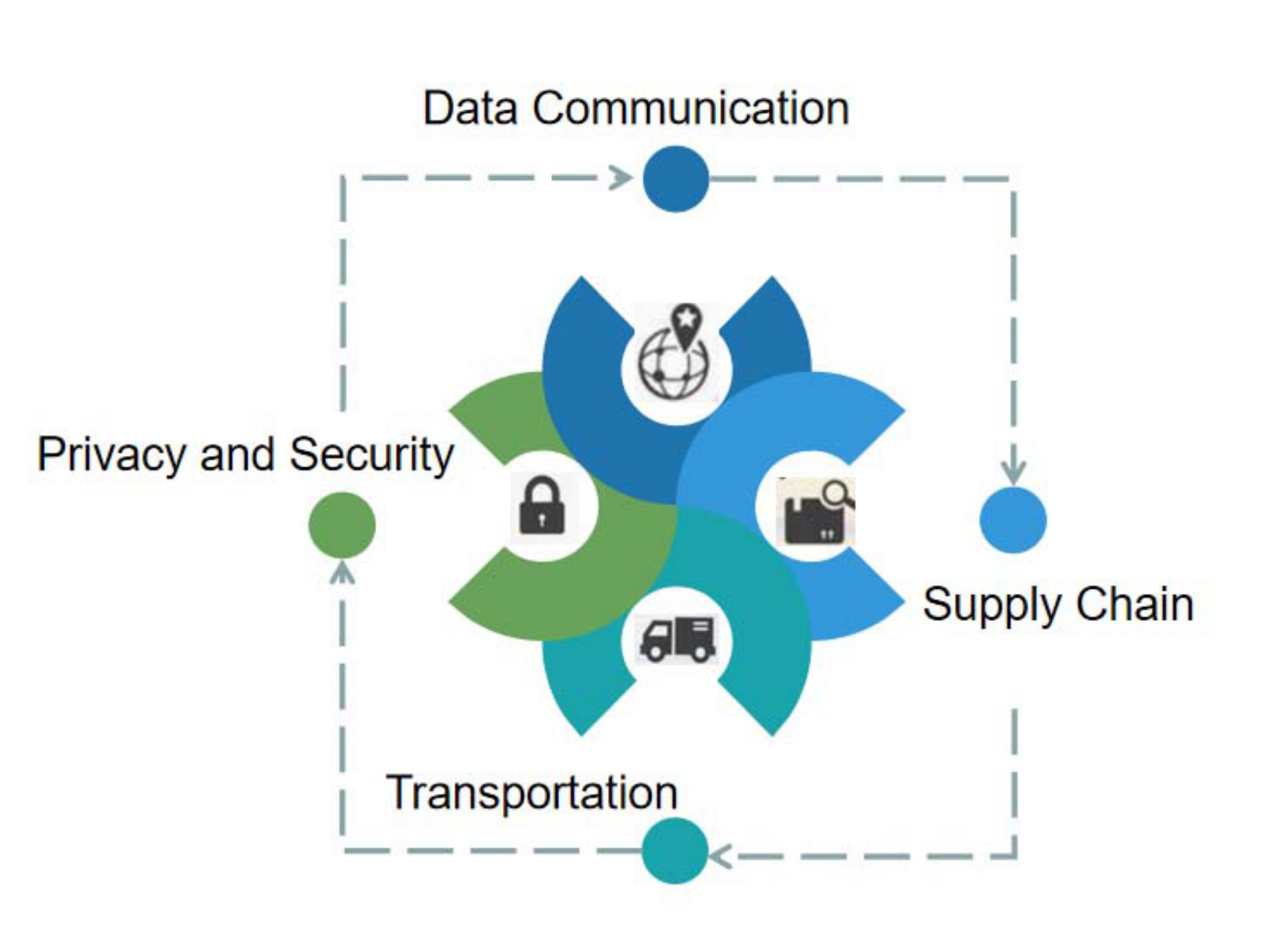}
\caption{Industrial logistics}
\label{logistics}
\end{figure}

\begin{table*}
\centering
\caption{Differential privacy for industrial logistics system}
\begin{tabular}{|m{0.6cm}|m{0.6cm}|m{3cm}|m{2.5cm}|m{2cm}|m{6.5cm}|}
\hline
\specialrule{0em}{2pt}{2pt}\makecell[c]{\textbf{Ref}} & \textbf{Year}  &  \textbf{Privacy Technology} & \textbf{Problem Solved}  &  \textbf{ Criterion} & \makecell[c]{\textbf{Contributions}} \\
\hline
\cite{Primault2019The} & 2018 & Comprehensive survey  & Location privacy  & Diversification & \makecell{Provide an up-to-date vision on location privacy \\ Organise LPPMs into three use cases } \\
\hline
\cite{Yin2018Location} & 2017 & Adjusted differential privacy & Location privacy  & $\varepsilon$-differential privacy  & \makecell{ LPT for representing location data  \\ DP-k model \\ Extensive experiments on real-world datasets } \\
\hline
\cite{2018DP3} & 2018 & DP3 scheme & Privacy-preserving indoor localization  & $\varepsilon$-differential privacy & \makecell{  Balance the trade-off between data privacy
and data utility\\ Online real-time operating phase } \\
\hline
\cite{0PAPU} & 2020 & PAPU for differential privacy & Protecting vehicles¡¯ trajectory privacy  & $\varepsilon$-differential privacy & \makecell{   Pseudonym swap
process based on differential privacy \\  New pseudonym swap mechanism \\ Guarantee the unlinkability  } \\
\hline
\cite{Wei2019Differential} & 2019 & DPLP scheme & Location protection in spatial crowdsourcing  &$\varepsilon$-differential privacy & \makecell{ Novel DP-based location protection \\  $\varepsilon _{1}$-ATGD and $\varepsilon _{2}$-DPACPG } \\
\hline
\cite{Zhang2018Enabling} & 2018 & Probabilistic differential privacy & Location recommendations  & $(\varepsilon, \sigma )$-differential privacy  & \makecell{ Investigate fine-grained location recommendations \\  Lower bound of the variety of aggregate statistics \\  Extensive experiments on accuracy, privacy and efficiency} \\
\hline
\cite{Li2020A} & 2020 & PPDP algorithm & Transit card data privacy & $\varepsilon$-differential privacy & \makecell{ New prefix tree structure \\ Incremental privacy allocation mechanism \\ Spatial-temporal domain reduction model } \\
\hline
\cite{Wang2019Equivalent} & 2019 & Equivalent mechanism & Releasing location data  & $(\varepsilon, \sigma )$-differential privacy  & \makecell{ Determine location errors on indistinguishability \\ Equivalent mechanism to enforce differential privacy } \\
\hline
\cite{Liu2018Trajectory} & 2018 & Spatial streaming data with differential privacy & Trajectory privacy protection  & $(\omega, n) $-differential privacy & \makecell{ Flexible trajectory privacy model of w-event n2-block \\ Spatial temporal budget distribution (STBD) algorithm } \\
\hline
\end{tabular}
\label{TABLogistics}
\end{table*}

\subsection{Industrial Logistics System}

Industrial logistics is the guarantee of timely and effective transportation system, which can help improve the efficiency of industrial production.
Generally, industrial logistics system has four pillars: 1) IT security and privacy; 2) communication system; 3) supply chain monitoring and 4) transportation tracking, as shown in Fig. \ref{logistics}. It is difficult to migrate between customers with different industry attributes, so industrial logistics focuses on matching autonomous mobile robot (AMR) technology with user value, and realizes intelligent enabling in complex environment.

In terms of longer development cycle, industrial logistics has changed from automation to intelligence, and the protection of logistics privacy data is highlighted \cite{cai2019differential}. Specially, differential privacy is worthy of attention. We summarized the important existing papers in Table \ref{TABLogistics}.

In the beginning, some research results focus on the privacy protection of logistics or location, instead of particular differential privacy. Primault \emph{et al.} \cite{Primault2019The} proposed a survey on computational location privacy, and authors focused on privacy threats in common IoT scenarios . It can be concluded that many essential contents of IIoT scenarios are the same as traditional ones.
Yin \emph{et al.} \cite{Yin2018Location}  studied this problem in the context of big data of IIoT, especially in low density of location data with high value. Authors built the multilevel location information tree model.

Some research results focus on the direct use of differential privacy.
Wang \emph{et al.}  \cite{2018DP3} proposed the DP3 for  privacy-preserving indoor localization mechanism based on differential privacy.
Li \emph{et al.}  \cite{0PAPU}  built the PAPU (Pseudonym Swap with Provable Unlinkability) for VANETs (Wireless communication technology, vehicle ad hoc networks) based on differential privacy.
Wei \emph{et al.} \cite{Wei2019Differential}  focused the location protection in spatial crowdsourcing  under the help of differential privacy.
Zhang and Chow \cite{Zhang2018Enabling}  proposed a method for location recommendations protection based on enabling probabilistic differential privacy.
Li \emph{et al.}  \cite{Li2020A}  solved the  smart card data transiting with privacy-preserving data publishing algorithm based on differential privacy.
Wang \emph{et al.}  \cite{Wang2019Equivalent} proposed the  equivalent mechanism, which can be used for releasing location data based on differential privacy.
Liu  \emph{et al.} \cite{Liu2018Trajectory}  achieved trajectory privacy protection designed for spatial streaming data under the help of differential privacy.

In addition to cloud and analysis, the IIoT is the driving force connecting the logistics field, and freight monitoring is one of the leading application scenarios. In industrial transportation system, logistics planning and location information involve data privacy.

\begin{table*}
\centering
\caption{Differential privacy for  power and energy: smart grid}
\begin{tabular}{|m{0.6cm}|m{0.6cm}|m{3cm}|m{2.5cm}|m{2cm}|m{6.5cm}|}
\hline
\specialrule{0em}{2pt}{2pt}\textbf{Ref} & \textbf{Year}  &  \textbf{Privacy Technology} & \textbf{Problem Solved}  &  \textbf{Criterion} & \makecell[c]{\textbf{Contributions} }\\
\hline
\cite{2019Towards} & 2019 & DP-based auction market  & Island MicroGrids  & $(\varepsilon, \sigma )$-differential privacy & \makecell{Novel online double auction scheme \\ Two-phase differential privacy \\Extensive performance evaluation } \\
\hline
\cite{Fioretto2019Differential} & 2019 & PLO-differential privacy & Power grid obfuscation & $\varepsilon$-differential privacy & \makecell{Power Line Obfuscation (PLO) mechanism \\ Strong theoretical properties \\ Handle time-series network data } \\
\hline
\cite{2020Singular}  & 2020 & SSA-Local differential privacy & Prevent inferring household appliance classification & $\varepsilon$-differential privacy  & \makecell{Singular spectrum analysis \\ Fourier spectral noise \\ Detailed theoretical analysis }    \\
\hline
\cite{2017Cost} & 2016& Battery-based differential privacy & Achieve DP and cost saving simultaneously & $(\varepsilon, \sigma )$-differential privacy & \makecell{ Battery-based DP-preserving \\ Cost-friendly DP-preserving schemes  \\ Detailed theoretical analysis }    \\
\hline
\cite{Lou2020Cost} & 2020  & Shapley cost sharing scheme & Demand reporting for smart grids & $\varepsilon$-differential privacy & \makecell{ Apply the principle of Shapley value \\ Analytic expression for the total privacy cost \\  Demand reporting scheme } \\
\hline
\cite{Hassan2019Differential} & 2019& DPLM algorithm & Renewable energy resources & $\varepsilon$-differential privacy & \makecell{ Preserve privacy of  RERs integrated smart meter users \\ Preserve information about intermittent availability \\ Develop an algorithm for monthly accumulation  }\\
\hline
\cite{Mak2020Privacy} & 2020 &  Privacy-preserving obfuscation & Distributed power systems   & $\varepsilon$-differential privacy & \makecell{  Novel and distributed PD-OPF mechanism \\ Experiments on large collection of OPF benchmarks   } \\
\hline
\cite{Wang2018Data} & 2018 & Distributed differential privacy  & Achieve DP and cost saving simultaneously &$(\varepsilon, \sigma )$-differential privacy & \makecell{   Datadriven  differential privacy\\  Formulate the cost minimization problem into a RASP   }\\
\hline
\cite{Zhao2014achieving} & 2014 & Randomized BLH and Multitasking-BLH-Exp3 & Prevent inferring household appliance information  & $(\varepsilon, \sigma )$-differential privacy & \makecell{ Investigate the privacy issues of the smart
meters \\ Utilize Exp3 algorithm  for MAB }
 \\
\hline
\end{tabular}
\label{TABSmartGrid}
\end{table*}

\subsection{Smart Grid}

In order to achieve smart city \cite{Song2017smart}, smart grid is the most important foundation. With the wide application of smart devices in smart grid, the relationship between power users and providers has become more and more close.  This two-way interaction ensures the real-time transmission of power consumption data. At the same time, the privacy problems of users are increasingly obvious, the privacy information leakage of power users is closely related to everyone, especially the performance, which is the limiting factor for the further development of smart grid. Therefore, it is of great significance to carry out the research on the privacy protection of power users in smart grid.

For the privacy issues of smart grid under the help of differential privacy, we summarized the important existing papers in Table \ref{TABSmartGrid}.

Li \emph{et al.}   proposed the   online double auction method for smart grid based on differential privacy \cite{2019Towards}.
Fioretto \emph{et al.} \cite{Fioretto2019Differential}   discussed the application of differential privacy in power grid obfuscation and proposed the Power Line Obfuscation(PLO).
Ou \emph{et al.}  \cite{2020Singular}  made a detailed singular spectrum analysis on smart grid classifications based on local differential privacy.
Zhang \emph{et al.}    \cite{2017Cost}  exploited the dual roles of noise and made use of cost-friendly differential privacy on smart meters.
Lou \emph{et al.} \cite{Lou2020Cost}  discussed the cost and pricing of differential privacy under demand reporting for smart grids.
Hassan \emph{et al.} \cite{Hassan2019Differential} proposed the renewable energy resources based smart metering based on  differential privacy .
Mak \emph{et al.} \cite{Mak2020Privacy} put forward the distributed power systems based on privacy-preserving obfuscation.
Wang \emph{et al.}  \cite{Wang2018Data}  designed a method to achieve data-driven optimization for utility providers based on differential privacy.
Zhao \emph{et al.}  \cite{Zhao2014achieving}  achieved  data disclosure in smart grid based on differential privacy.

\subsection{Industrial Bioengineering}

\begin{figure}
\centering
\includegraphics[width=3.5in,height=2in]{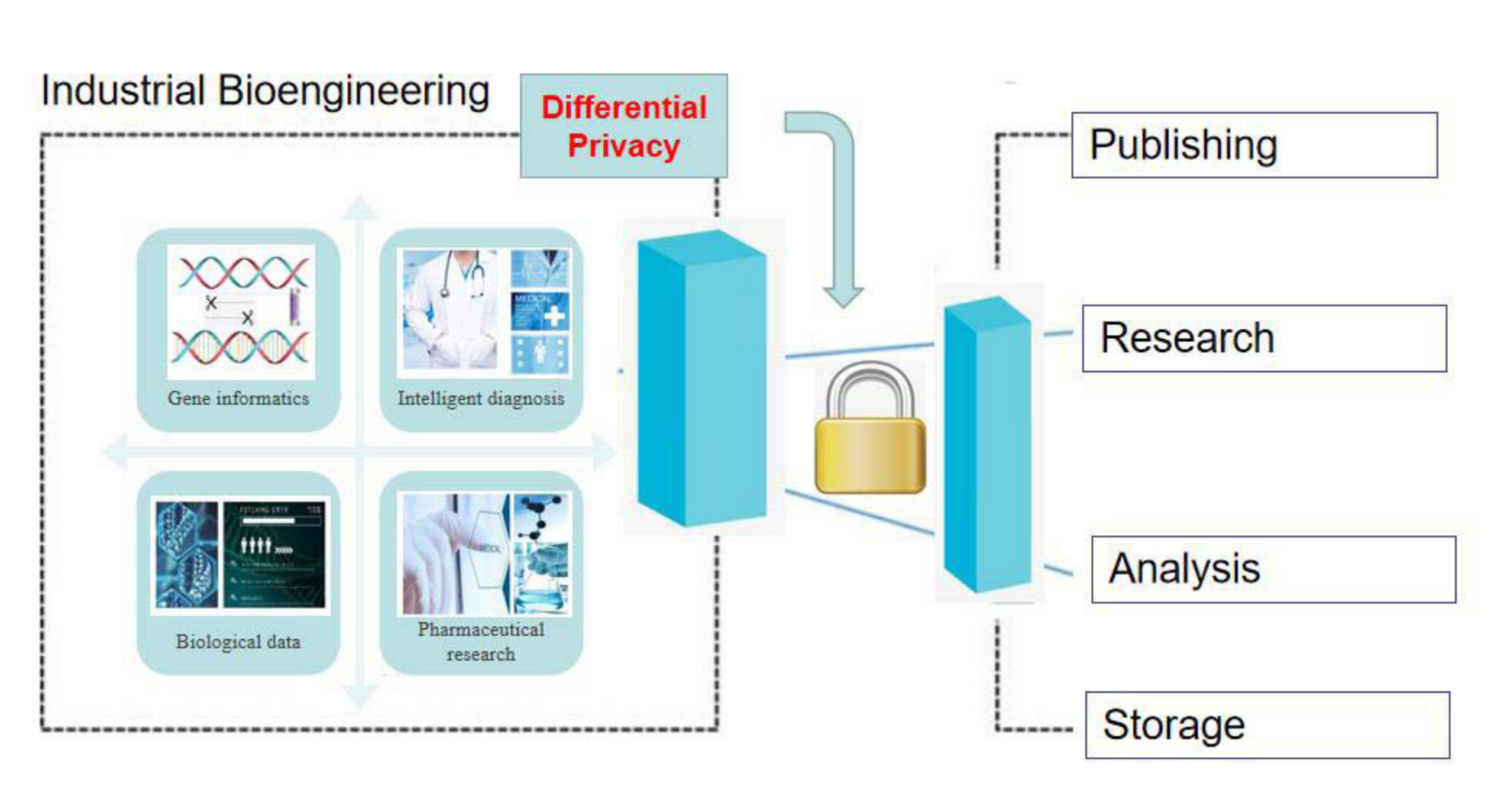}
\caption{Differential privacy for  bioengineering}
\label{fig04}
\end{figure}

In recent years, the industrial development of bioengineering has also attracted much attention, and the protection of personal sensitive data such as biometrics is extremely important, as shown in Fig. \ref{fig04}. Therefore, many researchers have begun to focus on the privacy protection of bio sensitive data, so as to lay the foundation for the further development of bio industry.

With the development of computer, optics, acoustics, biosensor and biostatistics, it is more and more common to use the inherent physiological characteristics of human body, such as fingerprint, face, iris, and behavioral characteristics, such as handwriting, voice, gait, etc. Fingerprint recognition and face recognition are the most widely used. And privacy protection in this process is also very important \cite{lin2016differential}. We summarized the important existing papers in Table \ref{TABBioengineering}.

Raisaro \emph{et al.}  \cite{2018Protecting} discussed the protecting privacy for genomic data in i2b2 (Informatics for Integrating Biology and the Bedside) under help of homomorphic encryption and differential privacy.
Liu \emph{et al.}  \cite{2020Blockchain} proposed a method for coronary heart disease diagnosis in mobile edge computing based on blockchain-enabled contextual online learning.
Wei \emph{et al.} \cite{Wei2020Diff} made use of differential privacy to achieve genetic matching for personalized medicine .
Wang \emph{et al.} \cite{Wang2018Secure} put forward a method for secure medical data collection based on local differential privacy .
Li \emph{et al.}  \cite{Li2015Efficient} explored the efficient e-health data release based on differential privacy with consistency guarantee .

\begin{table*}
\centering
\caption{Differential privacy for  bioengineering}
\begin{tabular}{|m{0.6cm}|m{0.6cm}|m{3cm}|m{2.5cm}|m{2.5cm}|m{6cm}|}
\hline
\specialrule{0em}{2pt}{2pt}\textbf{Ref} & \textbf{Year}  &  \textbf{Privacy Technology} & \textbf{Problem Solved}  &  \textbf{Criterion} & \makecell[c]{\textbf{Contributions} }\\
\hline
\cite{2018Protecting} & 2018 & Combine DP and homomorphic encryption  & Genomic data privacy  & $\varepsilon$-differential privacy &  \makecell{  Advanced privacy-enhancing technologies \\Build system by most widespread framework}  \\
\hline
\cite{2020Blockchain} & 2020 & Local differential privacy & Coronary heart disease diagnosis & $\varepsilon$-differential privacy  &  \makecell{Novel context-aware online learning algorithm  \\   Adaptively expanding tree structure \\ Adopt the local DP method} \\
\hline
\cite{Wei2020Diff} & 2020 & DPE algorithm with DPNM  &  Genetic matching in personalized medicine & $\varepsilon$-differential privacy &  \makecell{ DP-based genetic matching (DPGM) \\  DP-based
EIGENSTRAT (DPE) algorithm \\ DP-based Norder Markov (DPNM) algorithm}\\
\hline
\cite{Wang2018Secure} & 2018 &  Local differential privacy  & Secure medical data collection & $\varepsilon$-differential privacy  &  \makecell{ Secure medical data collection framework \\    Apply framework on synthetic data } \\
\hline
\cite{Li2015Efficient} & 2015 & Consistency guarantee under DP   & E-health data release   & $\varepsilon_{1}/\varepsilon_{2}$-differential privacy &  \makecell{ Design a new private partition algorithm \\ Apply constrained inference in post-processing stage}\\
\hline
\end{tabular}
\label{TABBioengineering}
\end{table*}

\subsection{Industrial Unmanned Aerial Vehicle}

UAV control technology also occupies an important position in IIoT \cite{Zhang2020Drone, Jiang2018Multimedia}. In some industrial scenarios, UAV data acquisition and user privacy are difficult to separate. Therefore, it is necessary to focus on how to protect UAV data protection methods in the IIoT, and differential privacy also contributes to this direction. We summarized the important existing papers in Table \ref{TABuav}.

Kim \emph{et al.} \cite{Kim2017On}  explored the  differential privacy-preserving movements issues for UAV. And they also   proposed the  UDiPP as a whole framework for DP-privacy preserving movements of UAV in the background of smart cities \cite{A2019}.
In addition, Wang \emph{et al.} \cite{Wang2020Learning}  put forward a method for aerial secure federated learning, which can be used  for UAV-assisted crowdsensing .

\begin{table*}
\centering
\caption{Differential privacy for  industrial UAV}
\begin{tabular}{|m{0.6cm}|m{0.6cm}|m{3cm}|m{2.5cm}|m{2cm}|m{6.5cm}|}
\hline
\specialrule{0em}{2pt}{2pt}\textbf{Ref} & \textbf{Year}  &  \textbf{Privacy Technology} & \textbf{Problem Solved}  &  \textbf{Criterion} & \makecell[c]{\textbf{Contributions} }\\
\hline
\cite{Kim2017On} & 2017 &   Privacy-preserving UAV framework   & Minimizing movements of UAVs &  - & \makecell{ Provide privacy-preserving movements of UAVs \\ New  UDiPP graph model}\\
\hline
\cite{A2019} & 2019 &  UDiPP  & privacy for UAV in Smart Cities &  $\varepsilon$-differential privacy & \makecell{  Support privacy preserving UAV movement strategies  \\Define a MintUDiPP problem \\ Creation of UDiPP graph   }\\
\hline
\cite{Wang2020Learning} & 2020 & Aerial secure federated learning & UAV-assisted crowdsensing & $\varepsilon$-differential privacy  & \makecell{Proposed SFAC as federated learning \\ Investigate a consortium blockchain network \\  Evaluated via extensive simulations}\\
\hline
\end{tabular}
\label{TABuav}
\end{table*}

\subsection{Industry 4.0: Intelligent Manufacturing}

\begin{figure}
\centering
\includegraphics[width=3.3in,height=2.2in]{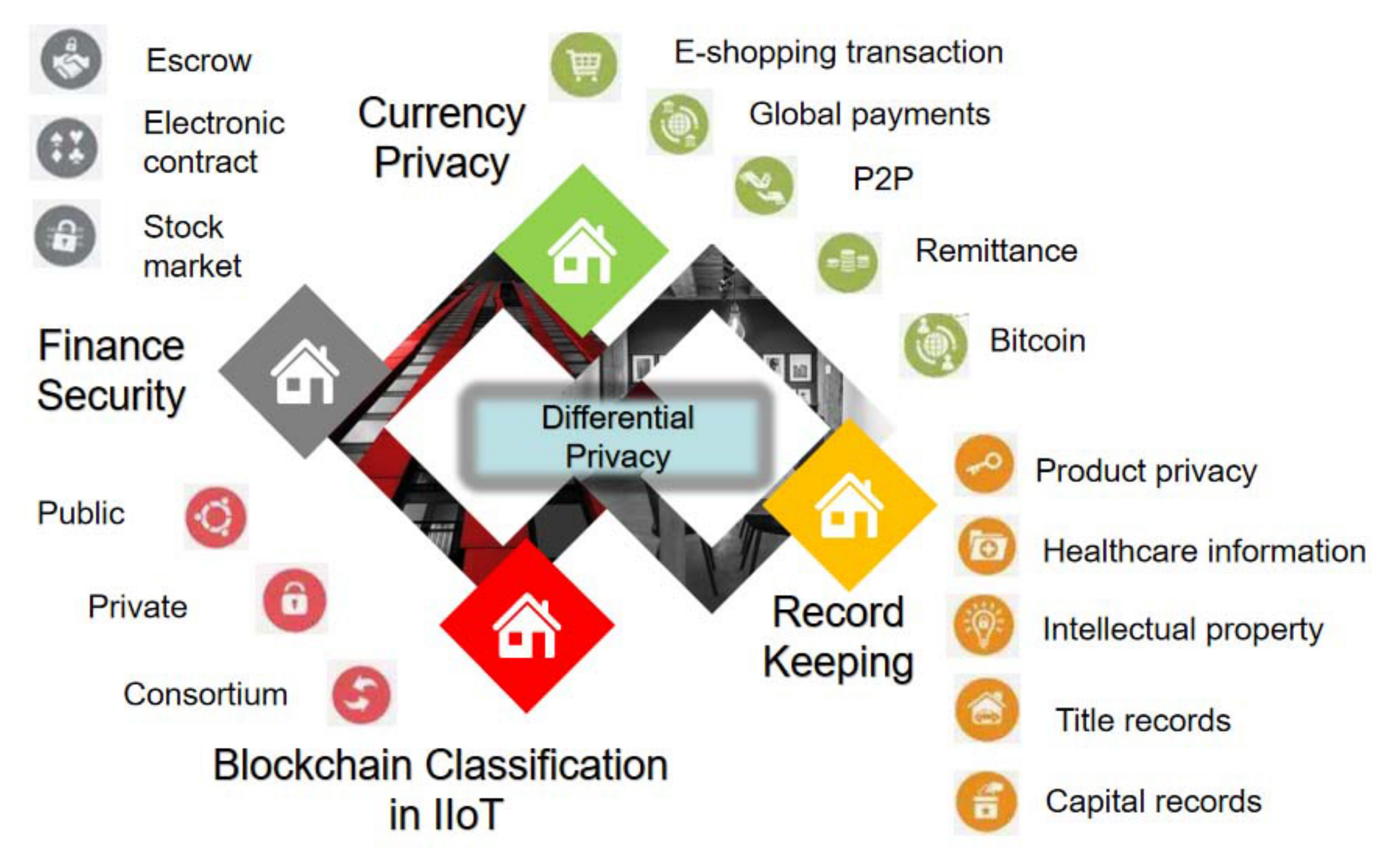}
\caption{Differential privacy for IIoT blockchain }
\label{in4}
\end{figure}

Intelligent manufacturing (IM) is a man-machine integrated intelligent system composed of intelligent machines and human experts. It can carry out intelligent activities in the manufacturing process, such as analysis, reasoning, judgment, conception and decision-making, as shown in Fig. \ref{in4}. Through the cooperation of human and intelligent machines, we can expand, extend and partially replace the mental work of human experts in the manufacturing process. It updates the concept of manufacturing automation and extends it to flexible, intelligent and highly integrated.

In fact, the privacy protection of industrial data is also very important. For example, the  automated guided vehicle (AGV) used in factory warehouse automation usually uses industrial WiFi for communication, and data leaks in this process are very common. Therefore, more and more researchers begin to discuss how to use differential privacy in IM.  We summarized the important existing papers in Table \ref{TABim}.

Wang \emph{et al.}  \cite{2017Differential}  discussed the  differential privacy for linear distributed control systems, which can be used for entropy minimizing mechanisms with performance tradeoffs.
Fan and Xiong \cite{2014An} proposed  an adaptive approach to realize the real-time aggregate monitoring based on adjusted differential privacy.
For sensor system, Chakraborty \emph{et al.}  \cite{Chakraborty2020Temporal} explored the  temporal differential privacy for industrial  wireless sensor networks. And  Wang \emph{et al.} \cite{Tian2020Edge} put forward   edge-based differential privacy computing method for industrial sensor-cloud systems. In addition, Barbosa \emph{et al.}  \cite{Barbosa2016A} proposed the technique to provide differential privacy for appliance usage in smart metering, which can be used for industrial intelligent privacy.

\begin{table*}
\centering
\caption{Differential privacy for  intelligent manufacturing}
\begin{tabular}{|m{0.6cm}|m{0.6cm}|m{3cm}|m{2.5cm}|m{2cm}|m{6.5cm}|}
\hline
\specialrule{0em}{2pt}{2pt}\textbf{Ref} & \textbf{Year}  &  \textbf{Privacy Technology} & \textbf{Problem Solved}  &  \textbf{Criterion} & \makecell[c]{\textbf{Contributions} }\\
\hline
\cite{2017Differential} & 2017 &  Entropy minimizing mechanisms for DP & Linear distributed control systems& $\varepsilon$-differential privacy  & \makecell{ Differential privacy of agents' preference vectors \\ Achieve DP by adding noise to shared information }\\
\hline
\cite{2014An} & 2013 & Adaptive approach based on DP & Real-time aggregate monitoring & $\varepsilon$-differential privacy & \makecell{ Establish the state-space model for the time series \\ Sample the time series data as needed \\ Formal analysis on filtering with fixed-rate sampling}\\
\hline
\cite{Chakraborty2020Temporal}& 2020 & Temporal differential privacy &  Wireless sensor networks & $\varepsilon$-differential privacy & \makecell{Temporal DP preserving mechanism \\ Time of occurrence is made indistinguishable}\\
\hline
\cite{Tian2020Edge}& 2020 & Edge-based differential privacy computing & Sensor-cloud systems & $\varepsilon$-differential privacy & \makecell{  Three-layer storage architecture \\ Research on the characteristics of raw data}\\
\hline
\cite{Barbosa2016A}& 2016 & Lightweight approach based on DP & Appliance usage in smart metering & $\varepsilon$-differential privacy & \makecell{ Measure privacy achieved by appliances \\ Evaluate  attack to eliminate  noise }\\
\hline
\end{tabular}
\label{TABim}
\end{table*}

\subsection{Blockchain in Industrial Informatics}

IIoT has profoundly changed the mode of production, organization and business model of traditional industries. Traditional technology has been unable to meet the needs of the future IIoT, but the blockchain technology provides trust, transparency and security communication guarantee for the IIoT with the characteristics of decentralization, openness, transparency and unforgeability \cite{Abdallah2020Blockchain}. For more basic content, Dai  \emph{et al.} \cite{Dai2019Blockchain}  made a comprehensive survey on  blockchain for IoT .

Blockchain technology can provide point-to-point direct interconnection for the IIoT to transmit data, rather than through the CPU, so that distributed computing can handle hundreds of millions of transactions.
At the same time, it can also make full use of the computing power, storage capacity and bandwidth of hundreds of millions of idle equipment distributed in different locations for transaction processing and greatly reduce the cost of calculation and storage \cite{Li2017Consortium}.
The security of blockchain in IIOT is more critical, and it puts forward new requirements for privacy protection technology. Blockchain technology superimposed with smart contract can turn each intelligent device into an independent network node that can maintain and adjust itself \cite{Liu2019Performance}. These nodes can exchange information with other nodes or verify their identity on the basis of pre-defined or embedded rules. We summarized the important existing papers in Table \ref{TABblock}.

Gai \emph{et al.}  \cite{Gai2019Privacy} designed a framework for privacy-preserving consortium blockchain for  energy trading , and it can help keep privacy  in industrial smart grid. In addition,
Yao \emph{et al.} \cite{2019Resource} proposed the method for resource trading in blockchain which can be used for IIoT.
The above papers focused on blockchain in IoT or IIoT, and there are also some researches focusing on differential privacy in blockchain directly.
Hassan  \emph{et al.} \cite{Hassan2020Differential}  made a detailed survey for differential privacy in blockchain technology, and it conducted the review with   a futuristic perspective.
For subtopics, Lu \emph{et al.} \cite{Lu2020Blockchain}  put forward a method based on blockchain with  federated learning, which can be used  for privacy-preserved data sharing in IIoT.
Gai  \emph{et al.} \cite{Gai2019Differentialg}  discussed the differential privacy-based blockchain technology for IIoT.
Roy \emph{et al.} \cite{Roy2020Blockchain} proposed the  blockchain-enabled safety-as-a-service for industrial IoT applications.

\begin{table*}
\centering
\caption{Differential privacy for  industrial blockchain }
\begin{tabular}{|m{0.6cm}|m{0.6cm}|m{3cm}|m{2.5cm}|m{2cm}|m{6.5cm}|}
\hline
\specialrule{0em}{2pt}{2pt}\textbf{Ref} & \textbf{Year}  &  \textbf{Privacy Technology} & \textbf{Problem Solved}  &  \textbf{Criterion} & \makecell[c]{\textbf{Contributions}} \\
\hline
\cite{Dai2019Blockchain}& 2019 &  Comprehensive survey & Blockchain privacy for IoT &  Diversification & \makecell{ Brief introduction on IoT and blockchain \\ Convergence of blockchain and IoT  \\ Blockchain for 5G-beyond networks in IoT}\\
\hline
\cite{Gai2019Privacy}& 2019 & Privacy-preserving consortium blockchain &  Energy trading   & $\varepsilon$-differential privacy &  \makecell{ Hide the trading distribution trends \\ Design  mechanism to introduce dummy accounts }\\
\hline
\cite{Lu2020Blockchain}& 2019  &   Blockchain and federated learning  & Data sharing in Industrial IoT & $\varepsilon$-differential privacy &  \makecell{ Transform  data sharing into machine learning problem  \\ New blockchain empowered collaborative architecture \\ Integrate differential privacy into federated learning}\\
\hline
\cite{2019Resource}& 2019 & IIoT DAO platform & Resource trading in blockchain for IIoT &  - & \makecell{  Use blockchain to construct decentralized trading platform \\ Model interaction between cloud provider and miners}\\
\hline
\cite{Gai2019Differentialg}& 2019 &  Blockchain-based Internet of edge model & Scalable and controllable IoT system & $\varepsilon$-differential privacy & \makecell{ Integrates IoT with edge computing and blockchain \\ Prevent data mining-based attacks}\\
\hline
\cite{Hassan2020Differential}& 2020 &   From futuristic perspective & DP in blockchain technology & Diversification &  \makecell{Provide importance of DP in blockchain \\ Presented future research directions}\\
\hline
\cite{Roy2020Blockchain}& 2020 &  Safety-as-a-service infrastructure    &  Prior intimation for safety-related information& - &  \makecell{ Blockchain integration into the Safe-aaS \\ Overall throughput follows an increasing trend}\\
\hline
\end{tabular}
\label{TABblock}
\end{table*}

\subsection{Industrial Social Network}

The social network model is also applicable in industrial IOT. In industrial IOT sensing, the information collected by each sensor can be regarded as personal information release of social network terminal. Privacy data is related to the quality of a production system.

In fact, social networks in industrial environments involve more production information and commercial information, and their privacy protection needs are more prominent than those of traditional social networks \cite{cai2016collective}.

The original intention of differential privacy is to protect the sensitive information of data release, which can be directly applied to the information collection of industrial IoT sensor, and help the industrial system to control the macro data such as production quality and product stability. We summarized the important existing papers in Table \ref{TABsocial}.

Abawajy \emph{et al.} \cite{Abawajy2016Privacy}  conducted the survey on data publication method for  privacy preserving social network.   Huang \emph{et al.}  \cite{Huang2020Privacy} put forward the  privacy-preserving approach PBCN for social network based on  differential privacy.
Hong \emph{et al.} \cite{Hong2015Collaborative} designed the  collaborative search log sanitization under help of differential privacy and boosted utility.
Wang \emph{et al.} \cite{2016Real} proposed method for protection crowd-sourced social network data  by enhanced RescueDP, especially for the real-time and spatio-temporal data.
Du \emph{et al.}  \cite{Du2017A}   put forward the  query model designed for sustainable fog data under the help of differential privacy.
Liu \emph{et al.} \cite{2018EPIC} defined the EPIC as a  Framework to weaken the Internet traffic analysis, which is also under help of differential privacy.
And Wang \emph{et al.} \cite{2019Differential} focused on   mobile social video prefetching based on  DP-oriented distributed online learning.
In addition, Lin \emph{et al.}  \cite{Lin2020Protecting} tried to  protect user's shopping preference based on differential privacy and it can help protect the social trading privacy.
Chamikara \emph{et al.} \cite{Chamikara2020Privacy} explored the related methods for  privacy preserving face recognition by differential privacy.
Liu \emph{et al.}  \cite{Liu2020Local}  discussed the  social network publishing based on local differential privacy and
Wei \emph{et al.} \cite{Wei2019Differentiall} focused the  differential privacy application in social network on trajectory community recommendation.

\begin{table*}
\centering
\caption{Differential privacy for  industrial social network}
\begin{tabular}{|m{0.6cm}|m{0.6cm}|m{3cm}|m{2.5cm}|m{2cm}|m{6.5cm}|}
\hline
\specialrule{0em}{2pt}{2pt}\textbf{Ref} & \textbf{Year}  &  \textbf{Privacy Technology} & \textbf{Problem Solved}  &  \textbf{Criterion} & \makecell[c]{\textbf{Contributions} }\\
\hline
\cite{Abawajy2016Privacy} & 2016 & Comprehensive survey  & Social network data publication& Diversification & \makecell{ High level social network threat analysis \\Categorize a spectrum of adversarial knowledge  \\ Graph structural-based privacy attack models }\\
\hline
\cite{Huang2020Privacy} & 2020 & DP based on clustering and noise & Social network privacy& $\varepsilon$-differential privacy & \makecell{ PBCN framework \\ Privacy measure by adjacency degree \\ Data sets with different sizes }\\
\hline
\cite{Hong2015Collaborative} & 2014 &  Differential privacy with boosted utility & Collaborative search log sanitization& $(\varepsilon, \sigma )$-differential privacy & \makecell{ Address deficiency by presenting sanitization \\ Prove differential privacy and protocol security for CELS  } \\
\hline
\cite{2016Real} & 2016 & Enhanced RescueDP  & Crowd-sourced social network data&  $\varepsilon$-differential privacy with w-event&\makecell{ Proposed RescueDP  \\ Enhanced RescueDP scheme\\  Evaluate method with real-world and synthetic datasets} \\
\hline
\cite{Du2017A} & 2017 & Differential privacy-based query model & Sustainable fog data centers & $(\varepsilon, \sigma )$-differential privacy &
\makecell{ Differential privacy-based query model \\ New query model for fog computing \\ QMA based on differential privacy  }\\
\hline
\cite{2018EPIC} & 2018 & EPIC: A Differential Privacy Framework & Preventing Internet traffic analysis & $\varepsilon d_{x}$-differential privacy & \makecell{DP mechanism for the selection of proxy gateways \\  DRW scheme for data transmissions  \\Simulations based on the real community topology} \\
\hline
\cite{2019Differential}& 2019 & DP oriented distributed online learning & Mobile social video prefetching& $\varepsilon$-differential privacy  &  \makecell{  Investigate relationship of user playback and demand  \\ Provide the privacy attacking model \\ Conduct  a series simulation tests}\\
\hline
\cite{Lin2020Protecting} & 2020 & Optimized differential private online transaction &   Protecting  shopping preference&  $(\varepsilon, \sigma )$-differential privacy &  \makecell{Protect consumption privacy in online banks \\The RO-DIOR scheme \\ The privacy loss is less than 0.5 }
\\
\hline
\cite{Chamikara2020Privacy} & 2020 & Adjusted privacy-preserving protocol   & Privacy preserving face recognition  & $\varepsilon$-differential privacy &  \makecell{ Privacy using eigenface perturbation \\ Towards controlled information release}  \\
\hline
\cite{Liu2020Local} & 2020 & Local differential privacy  & Social network publishing& $\varepsilon$-differential privacy  &  \makecell{ DP-LUSN\\  Implementation method for DP-LUSN\\Evaluate on three real-life social network datasets}\\
\hline
\cite{Wei2019Differentiall} & 2019 & DPTCR scheme  & Trajectory community recommendation in social network & $\varepsilon$-differential privacy & \makecell{ Novel DP-based trajectory community recommendation \\ Semantic expectation-based location transition algorithm }\\
\hline
\end{tabular}
\label{TABsocial}
\end{table*}

\section{Challenges of Differential Privacy in IIoT}
\label{sec7}

\begin{figure}
\centering
\includegraphics[width=3.1in,height=2in]{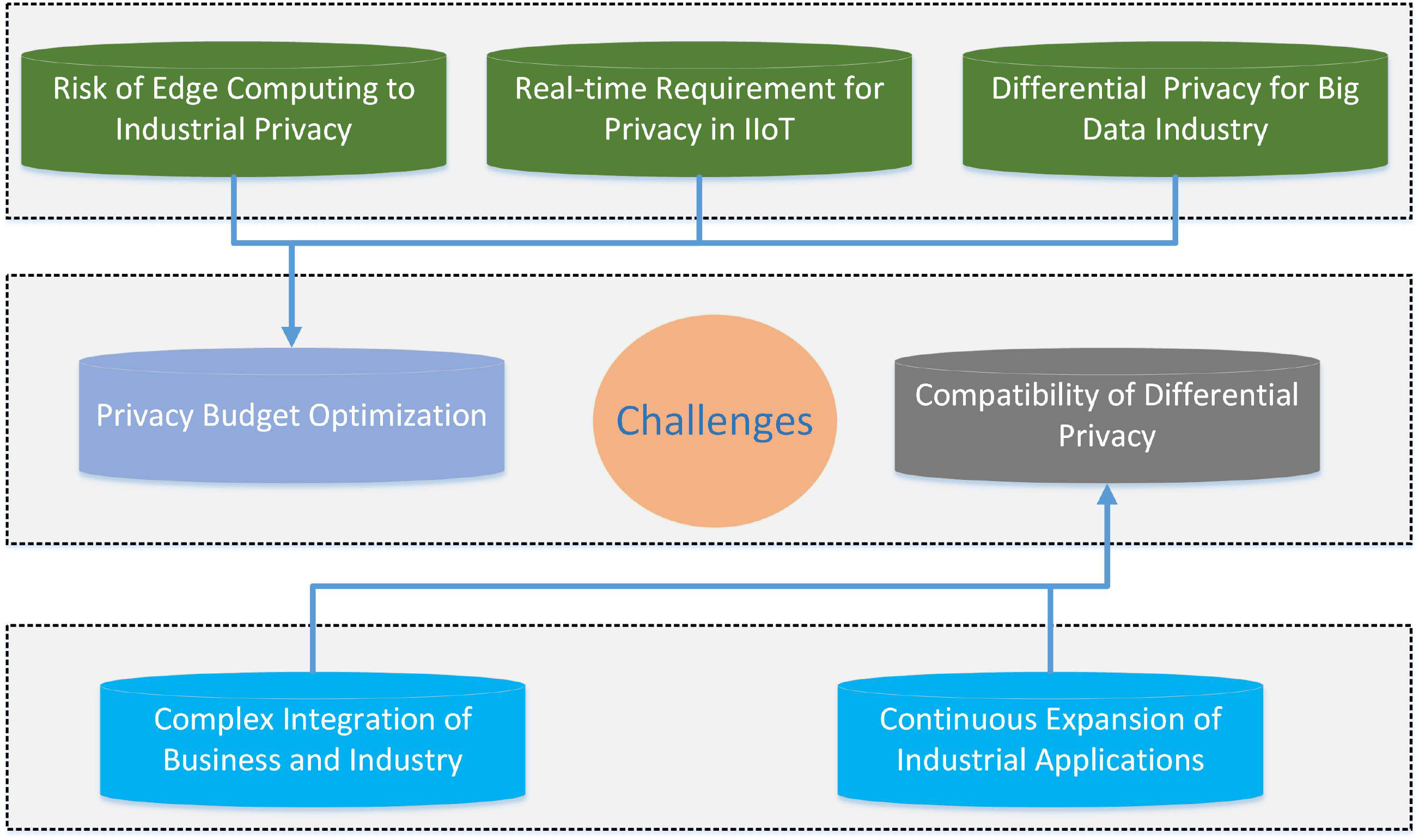}
\caption{Current challenges for DP in IIoT}
\label{challenge}
\end{figure}

At present, the development of differential privacy is still in a dynamic process, especially in the field of IIoT. In this survey, we summarize seven different problems in Fig. \ref{challenge}, which are urgent and meaningful. Specially, it contains the compatibility of differential privacy for industry, complex integration of business and industry,  differential  privacy for big data industry, risk of edge computing to industrial privacy, real-time requirement for privacy in IIoT, privacy budget optimization for privacy in IIoT and continuous expansion of industrial applications.

\subsection{Compatibility of Differential Privacy for Industry}

In initial stage,  differential privacy algorithm is designed for data publishing. Therefore, it has ushered an explosive development in the field of social and personal privacy information.
However, the types of industrial data and social information vary greatly, and the implementation means are different \cite{Lin2017A, Jiang2019Cyber}.

In this background, direct migration of differential privacy algorithm to industrial data will encounter a variety of complex contradictions \cite{2018Industrial}.
In the past, researchers usually directly migrate the differential privacy algorithm, which led to poor results.
Therefore, how to properly migrate and use differential privacy algorithm in industrial IoT is always one of the most important challenges.

\subsection{Complex Integration of Business and Industry}
\label{Integration}

The combination of business and industry brings more sensitive privacy data to industrial field.
And IIoT systems are also built based on integration of information technology and business process. For example, Mottola \emph{et al.} \cite{Mottola2018makeSense} conducted research on simplifying the integration of wireless sensor networks into business processes. And Culot \emph{et al.} \cite{Culot2019Integration} explored the  the integration and scale in the context of industry 4.0.

In addition, how to manage the balance between production and orders is an important problem for  operators. In order to solve it, more user data should be collected and used, which leads to more privacy threats.
Therefore, how to solve the privacy problem of IIoT under the commercial background has become a new problem.

\subsection{Differential  Privacy for Big Data Industry}

Industrial big data is a series of technologies and methods to excavate and display the value of data planning, collection, preprocessing, storage, analysis, mining, visualization and intelligent control \cite{Sun2016Internet}.
The essential goal of the research and breakthrough of industrial big data technology is to discover new patterns and knowledge from complex data sets, and to obtain valuable new information, so as to promote product innovation of manufacturing \cite{Chi2016A}. Big data means big risk. In  industrial environment, the use of big data is diverse, and the data sources are complex. So leakage risk of private data may be enhanced. Therefore, the application difficulty of differential privacy in this field is also obvious.

In this field, Zhou \emph{et al.} \cite{Zhou2019Differentially} discussed the multimedia big data retrieval based on edge computing under the differentially-private and trustworthy framework. However, the challenge of increasing the amount of data is very obvious. In addition,  D'Alconzo \emph{et al.} \cite{D'Alconzo2019A} conducted a survey on big data for network traffic monitoring and analysis, which also mentioned the privacy challenges brought by big data.

\subsection{Risk of Edge Computing to Industrial Privacy}

Industrial applications require more and more timely model updating. Due to the protection of sensitive data, some industrial operators such as factories are not willing to share data to the cloud.
As a solution, edge training plays a important role. In this way, what challenges need to be solved in order to achieve edge training and how can cloud edge work together better has been the focus of the industrial IoT.

For edge training in IIoT, some nodes provide data, and others are server nodes for training. So how to take differential privacy in this processing and how to divide the roles between nodes with suitable communication protocol are all the discussions under way. In order to solve this problem, Feng \emph{et al.}  proposed the privacy preserving high-order Bi-lanczos in fog computing, which is particularly designed for industrial applications \cite{Feng2020Privacy}.  In addition, Usman \emph{et al.}  \cite{Usman2020RaSEC} defined the RaSEC, which is an intelligent framework for reliable multi-level edge computing in industrial environments.

\subsection{Real-time Requirement for Privacy in IIoT}
The industrial field has high requirements for real-time and cooperative work with almost zero tolerance for delay. Langrica \emph{et al.} \cite{Langarica2020An}  explored the real-time fault diagnosis in industrial motors, which require fast computing IIoT.
Real time not only requires real-time computing, but also real-time data transmission.  Parizad \emph{et al.} \cite{Parizad2018Power}  discussed the power system real-time emulation, and proposed a practical virtual instrumentation.

In addition, industry is often a system rather than a single node running, requires the cooperation between the nodes.
It is inevitable to transmit and share data. Within a certain delay range, data can be guaranteed to arrive from one node to another, which is also the embodiment of real-time.
In this way, differential privacy algorithm design under real-time transmission is also big problem for IIoT.

\subsection{Privacy Budget Optimization for Privacy in IIoT}
For IIoT, differential privacy aims to strike a balance between data  and privacy, but it is not the end of the privacy puzzle.
Differential privacy relies privacy budget to adjust the privacy degree in system, which plays a decisive role in the effectiveness of privacy protection. Zhao \emph{et al.} \cite{Zhao2016Budget} designed a  budget-feasible incentive mechanisms for crowdsourcing tasks. And Han \emph{et al.} \cite{Han2019Differentially} proposed another differentially private mechanisms, which is also for budget limited mobile crowdsourcing.

In this field, how to define the rationality of privacy budget and how to control such privacy budget are still at the stage of research and exploration.  At the same time, many differential privacy algorithms used to generate noisy data depend on the industrial data to reach a certain size and meet a certain distribution, which may be difficult to meet in some specific IIoT scenarios. Especially when the added noise is too large, many application scenarios that provide personalized services based on personal information will encounter great challenges.

\subsection{Continuous Expansion of Industrial Applications}

At present, the achieved applications of IIoT mainly focus on intelligent industrial manufacturing, electric energy, automobile transportation, smart city and smart logistics.
With the continuous enrichment of industrial types,  ways for realizing of IIoT are expanding. So the requirements for privacy protection are also rising. In \cite{Serror2020Challenges}, Serror \emph{et al.} emphasized this problem.

In addition, the rapid development of AI has brought many new challenges to  differential privacy, and the intelligent attack and defense technology in this area is becoming one of the most urgent research hotspots.  Levesque \emph{et al.} \cite{Moren2018Growth} discussed the new type of IIoT, which can be defined as  growth through franchises in kowledge-intensive industries . These emerging IIoT Applications are bringing new challenges to differential privacy.

\section{Future Open Issues}
\label{sec8}

\begin{figure*}
\centering
\includegraphics[width=5.5in,height=3in]{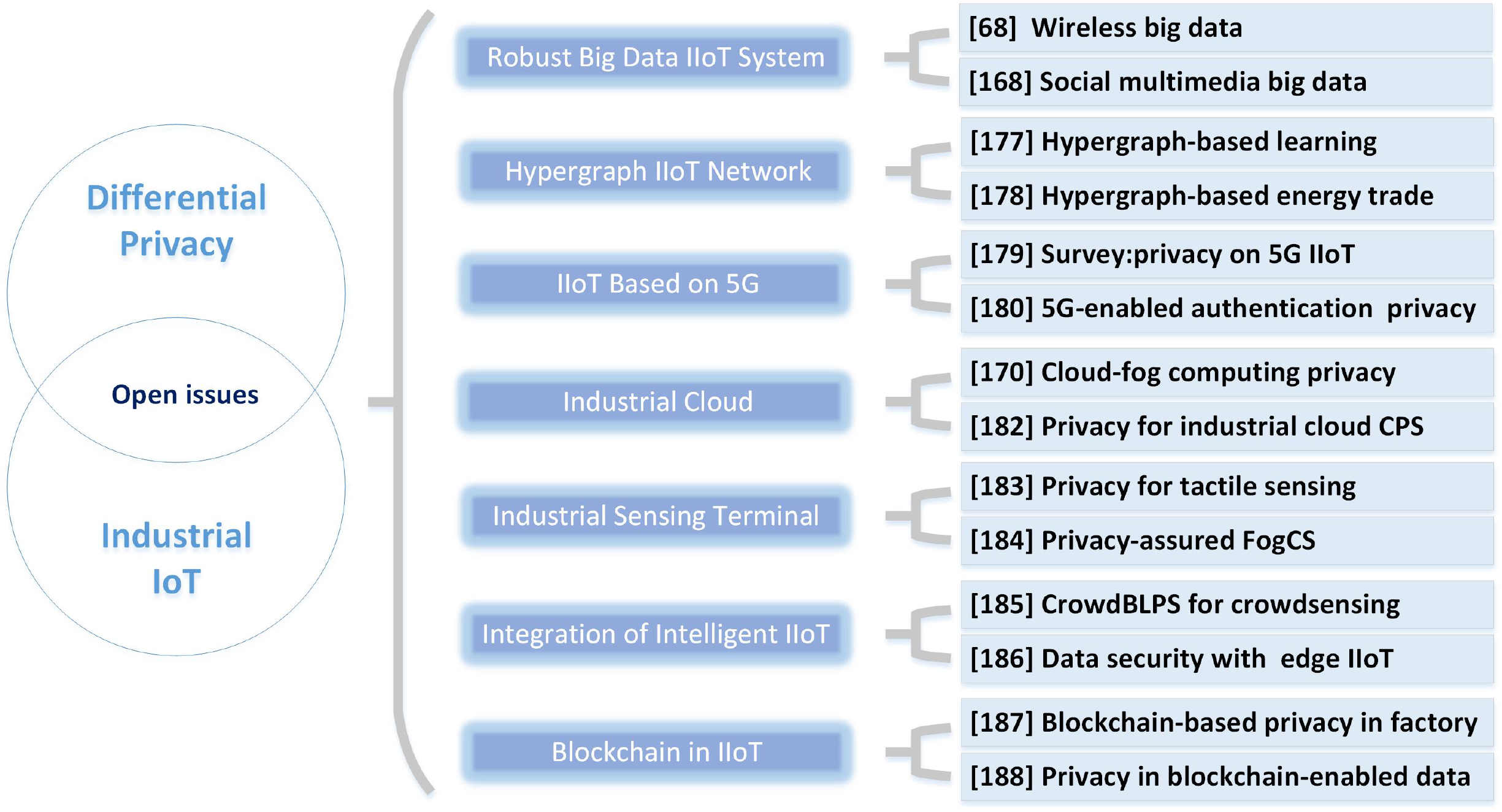}
\caption{Future open issues for  DP in IIoT}
\label{opportunities}
\end{figure*}

One of the main objectives in this survey is to propose more open ideas for research about differential privacy in IIoT.
Based on the analysis of the Section \ref{sec7}, several challenging problems have been summarized, and the hot directions are clear.
Here we put forward seven valuable subtopics in this field and discuss the feasibility of the related research one by one, as summarized in Fig. \ref{opportunities}.

\subsection{Big Data IIoT System with Robust Differential Privacy}

Industrial big data refers to the data generated in the application of informatization. With the deep integration of informatization and industrialization, information technology has penetrated into all aspects of the industrial chain of industry, such as QRC (Quick Response Code), RFID (Radio Frequency Identification), ERP (Enterprise Resource Planning) and so on. The expansion of data also provides a sufficient basis for AI algorithm, so as to launch more intelligent IIoT products. In this way,  it will be very valuable to study the differential privacy of industrial big data, especially based on differential privacy.

In this direction, some preliminary research results have been reported. Du \emph{et al.} \cite{Du2020Differential} explored the differential privacy preserving of training model under the background of  wireless big data.
Zhou \emph{et al.} \cite{Zhou2019Differentially} focused on  online social multimedia big data for privacy-preserving method.

\subsection{Differential Privacy for Hypergraph IIoT Network}

At present, the existing IIoT  information interaction technology has shortcomings in solving the problem of intelligent interaction, which is mainly manifested in failing to reflect the multidimensional, dynamic and complex network interaction relationship between IIoT entities, which can not meet the requirements of intelligent interaction.
According to the data characteristics of multidimensional dynamic mesh between entities, hypergraph theory can construct an entity relationship network model of IIoT.

In this context, it is necessary to propose more differential privacy algorithms for hypergraph IIoT network and achieve more valid protection for complex IIoT network. Although the problem has not been solved completely, some preliminary studies have been developed. Wang \emph{et al.}   \cite{Wang2020Security} proposed a directed hypergraph-based learning scheme based on security enhanced content sharing in social IoT. In addition, Karumba \emph{et al.}  \cite{Karumba2020HARB} defined the HARB, which can be regarded as  hypergraph-based adaptive consortium blockchain for decentralised energy trading.

\subsection{Differential Privacy of IIoT Based on 5G Communication}

In traditional communication technology, the main body of communication services is humans. For 5G Communication, situation will change.
It has been demonstrated that main users of communication services will gradually migrate from human to things, and the proportion will be higher and higher.
For IIoT, more industrial equipments can independently use communication tools to exchange information.

Because of this fundamental change, 5G will give birth to countless new fields, trigger a new round of industrial transformation, and become a new driver to promote IIoT development. In this background, the role of differential privacy is more important. It will provide a solid guarantee for the protection of equipment data privacy and keep maintain the safe operation of industrial systems. The combination of differential privacy and new wireless communication technology will become a new research hotspot. For example, Khan \emph{et al.}  \cite{Khan2019A} conducted a comprehensive survey on security and privacy of 5G. And Zhang \emph{et al.} \cite{Zhang2020Edge} proposed the edge computing-based  authentication framework considering privacy-preserving 5G-enabled vehicular networks.

\subsection{New Differential Privacy for Industrial Cloud  Data}

How to keep cloud data privacy without leaking  has become the main problem faced by the development of cloud computing, especially for IIoT.
It is particularly important to establish a set of security mode based on heterogeneous data, which can ensure data privacy in the system and application at the same time.

Edge computing and fog computing are often used interchangeably because they both involve pushing intelligence and processing power \cite{hu2017security}.
While there are clear differences between how and why to deploy which type of infrastructure, both are critical to a successful IIoT strategy.
In order to enable the future of IIoT, it is necessary to adopt the next generation solution including edge computing and fog computing in privacy, so as to expand the  devices, networks and applications.
Feng \emph{et al.}  \cite{Feng2020Privacy} discussed the privacy preserving problem in  cloud-fog computing for industrial applications.  Xu \emph{et al.} \cite{Xu2020PDM} put forward the privacy-aware deployment of machine learning for industrial cyber-physical could system.

\subsection{Differential Privacy  for Industrial Sensing Terminal}
As an important means to realize the comprehensive perception of IIoT, industry takes terminal sensors with various communication methods as the basic perception unit, in order to realize the distribution of perception tasks and the collection of sensing data. In this way, it can finally complete large-scale and complex  perception tasks for IIoT.

Swarm intelligence sensing applications need a large number of sensors to participate, and these data can carry sensitive information, making it face the risk of privacy disclosure. In this way, how to make use of differential privacy in IIoT sensing terminals is a hot topic for future research. In this filed, De Gregorio \emph{et al.} \cite{Daniele2018Integration} discussed the integration of robotic vision and tactile sensing for wire-terminal insertion tasks. Zhang \emph{et al.} \cite{Zhang2020Privacy} defined the privacy-assured FogCS, which can be used as chaotic compressive sensing for secure industrial big image data processing.

\subsection{Differential Privacy for Integration of Intelligent IIoT}

One of the ultimate goals  of IIoT is to realize the intelligent industrial production, so intelligence is also an important development trend for IIoT. With the introduction of AI platform, intelligent innovation in the field of IIoT will continue to emerge.

It has been demonstrated in Section \ref{Integration} that the integration poses more challenges for privacy protection in IIoT. As a main solution, the potential role of differential privacy is highlighted and has important research value. For example, Yu \emph{et al.}  \cite{Yu2019Toward} integrated the data security with  edge intelligent IIoT.  Zou \emph{et al.} \cite{Zou2020CrowdBLPS} proposed the  CrowdBLPS, which integrated the   blockchain with location-privacy-preserving for mobile crowdsensing system. On the whole, more integrations are needed to promote the further development of privacy technology.

\subsection{Differential Privacy for Blockchain in IIoT}

Most of the existing IIoT application systems take data transmission architecture in centralized and all terminals uniformly upload data to the cloud server. Under this architecture, the security and stability of cloud server is the key to the normal operation of the whole IIoT system. The application of blockchain in the industrial field can solve this problem.

Generally, the overall research of blockchain technology is still in hot spot, and its application in IIoT is more unique and has research value. How to directly or indirectly use differential privacy technology to protect the privacy information of blockchain will become the next research focus. Wan \emph{et al.} \cite{Wan2019A} proposed the blockchain-based solution designed for enhancing security and privacy in smart factory. And Liu \emph{et al.} \cite{Liu2019Blockchain} also discussed the privacy in blockchain-enabled data collection and sharing for IIoT with the help of reinforcement learning.

\section{Conclusions}
\label{sec9}

In this survey, we comprehensively reviewed the related researches on differential privacy in IIoT.
In order to provide more research ideas in the field of IIoT privacy protection, this survey presents an in-depth analysis of relevant topics.
Specially, we  reviewed related literature on IIoT and privacy protection, respectively. Then we focused on the metrics of industrial data privacy, and analyzed  contradiction between deep model data utilization and individual privacy protection. In current background, this survey also conducted detailed analysis of the opportunities, applications and challenges of differential privacy in IIoT.
Several valuable problems were identified and new research ideas were proposed in this survey.
It is hoped that this study can provide valuable references for researchers and promote the development of privacy protection of industrial IoT.

% or


\ifCLASSOPTIONcaptionsoff
  \newpage
\fi



%\bibliographystyle{IEEEtran}
%\begin{thebibliography}{1}


%\bibliographystyle{IEEEtran}
%\end{thebibliography}
\bibliography{bare_jrnl}


\bibliographystyle{IEEEtran}

\begin{IEEEbiography}[{\includegraphics[width=1in,height=1.25in,clip]{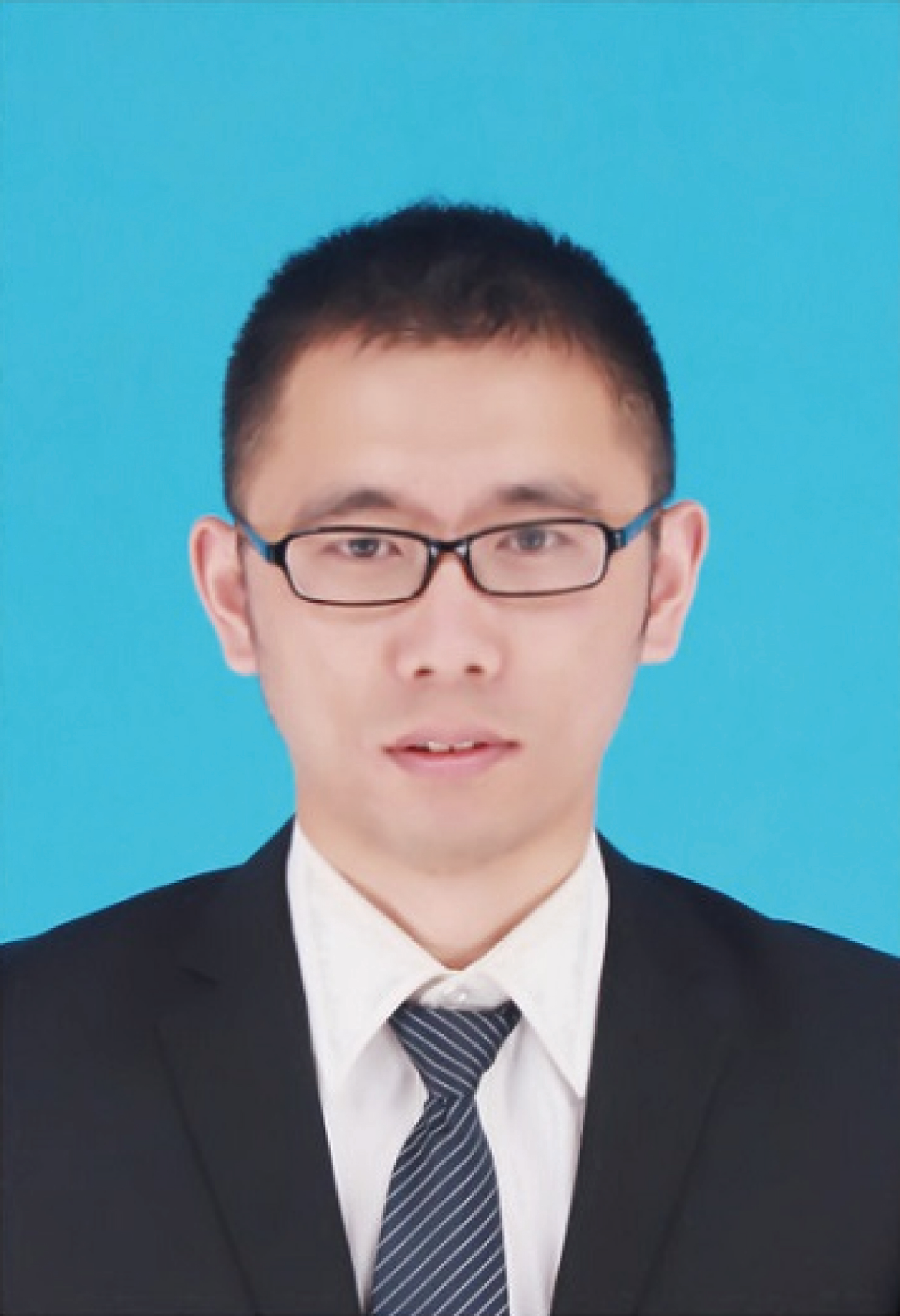}}]{Bin Jiang}
(Member, IEEE) received the B.S., M.S. and Ph.D. degrees in communication and information engineering from Tianjin University, China, in 2013, 2016, and 2020, respectively.
He was a visiting scholar in the Department of Electrical Engineering and Computer Science, Embry-Riddle Aeronautical University, Daytona Beach, FL, USA, from Nov.2017 to Feb.2019 and from Oct. 2018 to Oct. 2019.

Dr.Jiang is currently  a Postdoctoral Research Fellow in the College of Computer
and Software Engineering, Shenzhen University, Shenzhen, China, and he is also the member of the Security and Optimization for Networked Globe Laboratory, FL, USA.  He is the Editor for Frontiers in Communications and Networks, Guest Editor for Sensors, TCP Member for IEEE International Conference on Innovations in Information Technology and Program Committee Member for International Conference on Computer Engineering and Artificial Intelligence. His research interests lie in cybersecurity and privacy protection, industrial Internet of things and multimedia quality control.
\end{IEEEbiography}

\begin{IEEEbiography}[{\includegraphics[width=1in,height=1.25in,clip]{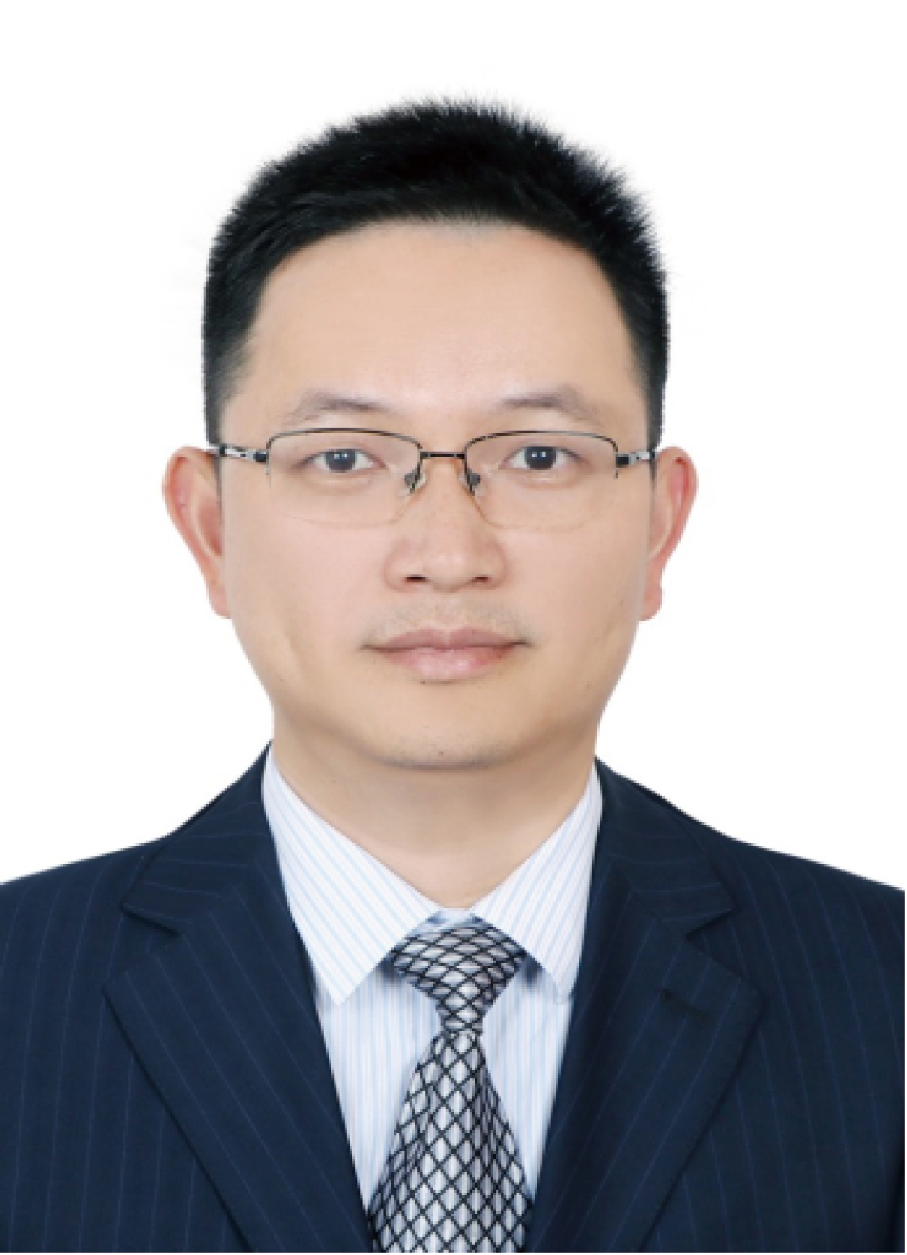}}]{Jianqiang Li}
received the B.S. and Ph.D. degrees
from the South China University of Technology,
Guangzhou, China, in 2003 and 2008, respectively.
He is a Professor with the College of Computer
and Software Engineering, Shenzhen University,
Shenzhen, China. He led three projects of the
National Natural Science Foundation and three
projects of the Natural Science Foundation of
Guangdong, China. His major research interests
include robotic, hybrid systems, Internet of Things,
and embedded systems.
\end{IEEEbiography}

\begin{IEEEbiography}[{\includegraphics[width=1in,height=1.25in,clip]{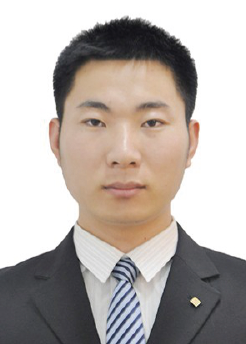}}]{Guanghui Yue}
received the B.S. and the Ph.D. degree in information and communication engineering from Tianjin University, Tianjin, China, in 2014 and 2019, respectively.
He was a joint Ph.D. student with the School of Computer Science and Engineering, Nanyang Technological University, Singapore, from 2017 to 2019. He is currently an Assistant Professor with the School of Biomedical Engineering, Health Science Center, Shenzhen University, Shenzhen, China. His research interests include bioelectrical signal processing and medical image analysis.
\end{IEEEbiography}

\begin{IEEEbiography}[{\includegraphics[width=1in,height=1.25in,clip]{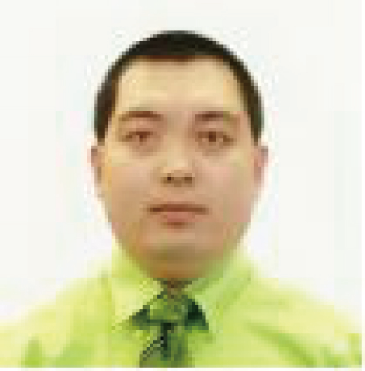}}]{Houbing Song}
(Senior Member, IEEE) received the Ph.D. degree in electrical engineering from the University of Virginia, Charlottesville, VA, in August 2012, and the M.S. degree in civil engineering from the University of Texas, El Paso, TX, in December 2006.

In August 2017, he joined the Department of Electrical Engineering and Computer Science, Embry-Riddle Aeronautical University, Daytona Beach, FL, where he is currently an Assistant Professor and the Director of the Security and Optimization for Networked Globe Laboratory (SONG Lab, www.SONGLab.us). He served on the faculty of West Virginia University from August 2012 to August 2017. In 2007 he was an Engineering Research Associate with the Texas AM Transportation Institute. He has served as an Associate Technical Editor for IEEE Communications Magazine (2017-present), an Associate Editor for IEEE Internet of Things Journal (2020-present) and a Guest Editor for IEEE Journal on Selected Areas in Communications (J-SAC), IEEE Internet of Things Journal, IEEE Transactions on Industrial Informatics, IEEE Sensors Journal, IEEE Transactions on Intelligent Transportation Systems, and IEEE Network. He is the editor of six books, including Big Data Analytics for Cyber-Physical Systems: Machine Learning for the Internet of Things, Elsevier, 2019,  Smart Cities: Foundations, Principles and Applications, Hoboken, NJ: Wiley, 2017, Security and Privacy in Cyber-Physical Systems: Foundations, Principles and Applications, Chichester, UK: Wiley-IEEE Press, 2017, Cyber-Physical Systems: Foundations, Principles and Applications, Boston, MA: Academic Press, 2016, and Industrial Internet of Things: Cybermanufacturing Systems, Cham, Switzerland: Springer, 2016.  He is the author of more than 100 articles. His research interests include cyber-physical systems, cybersecurity and privacy, internet of things, edge computing, AI/machine learning, big data analytics, unmanned aircraft systems, connected vehicle, smart and connected health, and wireless communications and networking. His research has been featured by popular news media outlets, including IEEE GlobalSpec's Engineering360, USA Today, U.S. News and World Report, Fox News, Association for Unmanned Vehicle Systems International (AUVSI), Forbes, WFTV, and New Atlas.

Dr. Song is a senior member of ACM and an ACM Distinguished Speaker. Dr. Song was a recipient of the Best Paper Award from the 12th IEEE International Conference on Cyber, Physical and Social Computing (CPSCom-2019), the Best Paper Award from the 2nd IEEE International Conference on Industrial Internet (ICII 2019), the Best Paper Award from the 19th Integrated Communication, Navigation and Surveillance technologies (ICNS 2019) Conference, the Best Paper Award from the 6th IEEE International Conference on Cloud and Big Data Computing (CBDCom 2020), and the Best Paper Award from the 15th International Conference on Wireless Algorithms, Systems, and Applications (WASA 2020).
\end{IEEEbiography}


%\vfill


\end{document}